\documentclass{JHEP3}

\usepackage{axodraw}
\usepackage{graphics}
\usepackage{epsfig}
\usepackage[active]{srcltx}

  \newcommand{\Mvariable}{}
  
  \newcommand{\Mfunction}{}

  \newcommand{\fract}[2]{{\textstyle \frac{#1}{#2}}}

  \newcommand{\ssh}{\not\!}
  \newcommand{\dd}{{\rm d}}
  
  \newcommand{\ima}{\mathop{\Im{\rm m}}}

\newcommand{\beq}{\begin{equation}}
\newcommand{\eeq}{\end{equation}}
\newcommand{\bea}{\begin{eqnarray}}
\newcommand{\eea}{\end{eqnarray}}
 \newcommand{\eqref}[1]{(\ref{#1})}
 \renewcommand{\epsilon}{\varepsilon}
 \renewcommand{\theta}{\vartheta}
 \renewcommand{\phi}{\varphi}

\newcommand{\ScaledGlueLoop}[4]{
   { \SetScale{#1}
  \SetScaledOffset(#2,#3)
  \begin{picture}(130,80)(-65,-40)
    \ArrowLine(-60,-35)(-50,0)
    \ArrowLine(-50,0)(-60,35)
    \Gluon(-50,0)(-20,0){5.0}{3}
    \Vertex(-20,0){1}
    \GlueArc(0,0)(20,0,180){5.0}{5}
    \GlueArc(0,0)(20,180,360){5.0}{5}
    \Vertex(20,0){1}
    \Gluon(20,0)(50,0){5.0}{3}
    \ArrowLine(60,-35)(50,0)
    \ArrowLine(50,0)(60,35)
    \Text(-1,-45)[]{#4}
  \end{picture}}}

\newcommand{\ScaledTadpole}[4]{
   {
   \SetScale{#1}
  \SetScaledOffset(#2,#3)
  \begin{picture}(130,80)(-65,-40)
    \ArrowLine(-60,-35)(-50,0)
    \ArrowLine(-50,0)(-60,35)
    \Gluon(-50,0)(50,0){5.0}{6}
    \GlueArc(0,18)(15,-90,270){5.0}{6}
    \ArrowLine(60,-35)(50,0)
    \ArrowLine(50,0)(60,35)
     \Text(-1,-45)[]{#4}
  \end{picture}}}

\newcommand{\ScaledGhostLoop}[4]{
   {\SetScale{#1}
  \SetScaledOffset(#2,#3)
  \begin{picture}(130,80)(-65,-40)
    \ArrowLine(-60,-35)(-50,0)
    \ArrowLine(-50,0)(-60,35)
    \Gluon(-50,0)(-20,0){5.0}{3}
    \Vertex(-20,0){1}
    \DashArrowArc(0,0)(20,0,180){5}
    \DashArrowArc(0,0)(20,180,360){5}
    \Vertex(20,0){1}
    \Gluon(20,0)(50,0){5.0}{3}
    \ArrowLine(60,-35)(50,0)
    \ArrowLine(50,0)(60,35)
    \Text(-1,-45)[]{#4}
  \end{picture}}}

\newcommand{\ScaledPinchedTriGraph}[4]{
   {\SetScale{#1}
  \SetScaledOffset(#2,#3)
  \begin{picture}(130,80)(-65,-40)
    \ArrowLine(-60,-35)(-50,0)
    \ArrowLine(-50,0)(-60,35)
    \GlueArc(-30,0)(20,0,180){5.0}{6}
    \GlueArc(-30,0)(20,180,360){5.0}{6}
    \Vertex(-50,0){5}
    \Vertex(-10,0){1}
    \Gluon(-10,0)(50,0){5.0}{5}
    \ArrowLine(60,-35)(50,0)
    \ArrowLine(50,0)(60,35)
\Text(-65,-36)[]{#4}
  \end{picture}}}

\newcommand{\ScaledBoxGraph}[3]{
   {\SetScale{#1}
  \SetScaledOffset(#2,#3)
  \begin{picture}(130,80)(-65,-40)
    \ArrowLine(-60,-35)(-30,-20)
    \ArrowLine(-30,-20)(-30,20)
    \ArrowLine(-30,20)(-60,35)
    \Gluon(-30,-20)(30,-20){5.0}{5}
    \Gluon(-30,20)(30,20){5.0}{5}
    \ArrowLine(60,-35)(30,-20)
    \ArrowLine(30,-20)(30,20)
    \ArrowLine(30,20)(60,35)
  \end{picture}}}

\newcommand{\ScaledPinchedBoxGraph}[4]{
   {\SetScale{#1}
  \SetScaledOffset(#2,#3)
  \begin{picture}(130,80)(-65,-40)
    \ArrowLine(-60,-35)(-25,0)
    \ArrowLine(-25,0)(-60,35)
    \GlueArc(0,0)(25,0,180){5.0}{6}
    \GlueArc(0,0)(25,180,360){5.0}{6}
    \Vertex(-25,0){5}
    \Vertex(25,0){5}
    \ArrowLine(60,-35)(25,0)
    \ArrowLine(25,0)(60,35)
  \Text(-65,-36)[]{#4}
  \end{picture}}}

\newcommand{\ScaledVertexGraph}[3]{
   {\SetScale{#1}
   \SetScaledOffset(#2,#3)
  \begin{picture}(130,80)(-65,-40)
    \ArrowLine(-60,-35)(-40,-20)
    \Gluon(-40,-20)(-40,20){5.0}{4}
    \ArrowLine(-40,20)(-60,35)
    \ArrowLine(-40,-20)(-10,0)
    \ArrowLine(-10,0)(-40,20)
    \Vertex(-10,0){1}
    \Gluon(-10,0)(40,0){5.0}{5}
    \ArrowLine(60,-35)(40,0)
    \ArrowLine(40,0)(60,35)
  \end{picture}}}

\newcommand{\ScaledPinchedVertexGraph}[4]{
   {\SetScale{#1}
   \SetScaledOffset(#2,#3)
  \begin{picture}(130,80)(-65,-40)
    \ArrowLine(-30,-35)(-20,0)
    \ArrowLine(-30,+35)(-20,0)
    \GlueArc(-38,0)(15,0,360){5.0}{8}
    \Gluon(-20,0)(50,0){5.0}{8}
    \ArrowLine(60,-35)(50,0)
    \ArrowLine(50,0)(60,35)
    \Vertex(-20,0){5}
    \Text(-65,-36)[]{#4}
  \end{picture}}}

\title{Gauge-Invariant Resummation Formalism and Unitarity in Non-Commutative QED}
\author{Nicola Caporaso\\
Dipartimento di Fisica, Polo Scientifico Universit\`a di Firenze,
INFN Sezione di Firenze\\
Via  G. Sansone 1, 50019 Sesto Fiorentino, Italy\\
Email: \email{caporaso@fi.infn.it}}
\author{ Sara Pasquetti \\
Dipartimento di  Fisica, Universit\`a  di Parma,
INFN Gruppo Collegato di Parma\\
Parco Area delle Scienze 7/A, 43100 Parma, Italy\\
E-mail: \email{pasquetti@fis.unipr.it}} \received{\today} 

\preprint{     
\hepth{0511127} \\ \today}    

\date{data}

\abstract{ We re-examine the perturbative properties
of four-dimensional non-commutative QED by extending the pinch
techniques to the $\theta$-deformed case. The explicit
independence of the pinched gluon self-energy from gauge-fixing
parameters and the absence of unphysical thresholds in the
resummed propagators permits a complete check of the optical
theorem for the off-shell two-point function. The known anomalous
(tachyonic) dispersion relations are recovered within this
framework, as well as their improved version in the (softly-broken)
SUSY case. These applications should be considered as a first step
in constructing gauge-invariant truncations of the
Schwinger--Dyson equations in the non-commutative case. An
interesting result of our formalism appears when considering the
theory in two dimensions: we observe a finite gauge-invariant
contribution to the photon mass because of a novel incarnation of
IR/UV mixing, which survives  the commutative limit when matter is
present. }

\keywords{Unitarity, Pinch-Techniques, Non-Commutative Gauge Theories}

\begin{document}

\section{Introduction}

The idea of introducing non-commutative space-time coordinates is
not new \cite{Snyder47} and has proved itself useful or
interesting in a wide range of different fields. Theoretical
high-energy physics observed a renewed interest toward
non-commutativity in the last few years, due to its relation with
string theory: the use of non-commutative geometry in this context
was pioneered by Witten \cite{Witten} in his formulation of open
string field theory. More recently compactifications of M-theory
on non-commutative tori were also studied in \cite{dh,cds}.
Finally, after the discovery that spacelike non-commutativity
emerges as an effective description of open strings in a constant
NS-NS $B_{\mu\nu}$ field \cite{SW}, this research line became
really fashionable. String theory reduces in a
particular low-energy limit to a quantum field theory on
non-commutative Minkowski space-time characterized by the algebra
\beq \label{ouralgebra} [\hat{x}_{\mu},\hat x_\nu] =
 i\theta_{\mu\nu}:
\eeq 
this fact generated a flurry of activity (see \cite{rev} for reviews)
to unveil the quantum properties of this novel class of models.
Unfortunately the commutation relations (\ref{ouralgebra}) entail
a breaking of Lorentz invariance, and difficulties arise at the
quantum level in obtaining the commutative $\theta\rightarrow 0$
limit in a sound way: These features are in open conflict with
observations, making the phenomenological prospects of
non-commutative models in particle physics quite thin \cite{banks,CCL1,CCL2}.

Nevertheless non-commutative QFT is very interesting in its own
right, presenting peculiar non-local interactions,
non-perturbative solutions \cite{insta} and unconventional
symmetries \cite{morita} that retain some properties of their
string and D-brane ancestors. Concerning the quantum consistency
of the theory, the loss of Lorentz invariance is not by itself a
catastrophe. While important modifications to the dynamics, like
non-trivial dispersion relations, may be introduced by breaking
Lorentz symmetry, most of the fundamental aspects of relativistic
QFT are retained, like microcausality, the CPT theorem, and so on
\cite{cpt}. Non-locality could instead drastically change the
quantum dynamics. A non-commutative action can be constructed by
deforming the ordinary, pointwise product of functions into the
Moyal star-product ($\tilde{f}(k)$ is the Fourier transformed
function)
\begin{equation}  \label{eq:star2}
f(x)\star\, g(x)= \int\frac{d^d k}{(2\pi)^d}\frac{d^d
k'}{(2\pi)^d}\,\tilde{f}(k)\,\tilde{g}(k-k')
\,e^{-\frac{i}{2}\theta^{\mu\nu}k_\mu k'_\nu}e^{i\,k'_\rho x^\rho}
\end{equation}
which manifestly induces terms with an arbitrarily high order of
derivatives in the action. This in turn implies an odd (IR/UV)
``mixing" of short and long distance scales by which, at the
quantum level, ultraviolet divergences are transferred to the
infrared domain \cite{MSV}: this effect impairs the familiar
Wilsonian point of view on renormalization \cite{MSV,KV,GP}.
Scalar theories have been widely studied and progresses have been
recently reported in constructing a renormalizable perturbative
expansion \cite{grosse}. The case of a gauge theory is more
difficult and no attempt has been done to prove systematically its
consistency: a serious conceptual obstacle appears because
non-commutative IR divergences induce tachyons at one-loop, and
these destabilize the perturbative vacuum unless additional matter
is introduced in a suitable way \cite{taki}. The relation between
these tachyonic instabilities and string theory dynamics has been
explored in \cite{adi}.

Vacuum destabilization leaves one to ponder if a stable vacuum
exists at all and, if this is the case, whether the breaking of
Lorentz invariance might make it possible for the theory to
develop exotic phases. These issues are intrinsically
non-perturbative, and it would be natural to take advantage of the
discretized formulation of non-commutative gauge theories
\cite{amb,Steinacker:2003sd,Grosse:2004wm,Behr:2005wp}. Insisting on a continuum description, two approaches
come instead to the mind: to write down an effective (CJT) action
for composite operators \cite{CJT}, and making use of the
Schwinger--Dyson equations (SD). Both approaches have been
extensively exploited in the commutative framework along the
years, and extensions to the non-commutative setup have been
accomplished for the $\lambda\phi^{\star 4}$ theory, suggesting
the possibility for a transition toward
 ``striped phases''
\cite{gubso}. The original proposal of the existence of a new
vacuum state breaking of translational invariance has been further
confirmed by analytical computations
\cite{Mandanici:2003vt,Castorina} and by numerical simulations (in
the lattice approach) \cite{latt}. Both the CJT effective action
and the gap equations obtained from the SD equations encode a
resummation of some infinite subset of Feynman diagrams.
Unfortunately, in a gauge theory the mechanism by which the
unphysical degrees of freedom cancel against each other calls into
action a large number of different Feynman diagrams, so that a
casual resummation of these is almost certain to waste
gauge-invariance, and yield a gauge-dependent answer to an
ostensibly gauge-independent question. Already in the commutative
case, for example, the SD make up a set of coupled non-linear
equations and a truncation (either on the number of loops, or on
the ``order'' of the $n$-point function examined, or both) is
necessary to obtain a tractable gap equation.  A ``gauge-invariant
resummable" formalism has been proposed \cite{lots} to achieve
such a truncation without introducing gauge artifacts. These
\textit{pinch techniques} (PT) consist in composing the ghost and
gauge-fixing-dependent degrees of freedom from different Feynman
diagrams in such a way that gauge-independent propagators and
vertices are defined \textit{before} the Schwinger--Dyson
equations are written down, making a gauge-independent truncation
possible.

The pinch techniques are by now well established in the framework
of ordinary gauge theories. They have been used to investigate the
generation of an effective gluon mass in QCD \cite{Cornwall}, to
properly describe resonant transition amplitudes and instable
particles \cite{resona}, \cite{PT2} and in QFT at finite temperatures they are
used to describe magnetic screening \cite{finitT}, to name just a
few applications. Computations have been carried out explicitly to
two-loops order \cite{pila} and in general gauges, covariant and
not, showing the consistency and unicity of the definition of
gauge-invariant propagators and vertices.

The main goal of this paper is the extension of the pinch
techniques to the non-commutative setup; the possibility of
writing down a gap equation is left for future work. A simpler
application which we will discuss here is the analysis of the
unitarity of the theory. For scalar field theories with purely
spacelike non-commutativity no loss of unitarity appears, but one
observes its violation in the timelike case (that is, when
$\theta^{\mu\nu} q_\nu$ has non-vanishing timelike components) in
the guise of unphysical imaginary parts in the particle's
self-energy \cite{GM,ABZ}. Actually it has been claimed that the
conventional perturbation theory is not suitable when
non-commutativity involves time and that a modified Dyson series
should be employed \cite{BDFP}. Unitarity has been checked within
this framework \cite{SL}, even though the relation of this
approach with the usual perturbative expansion and with the string
theory results seems to be mysterious.

Similar analyses have been attempted for non-commutative QCD
\cite{BGNV,BDF}. While in the standard formulation the timelike
case is confirmed to violate the optical theorem, unitarity of
spacelike non-commutative  gauge theories has only been checked for on-shell
propagators and/or in specific gauges. In particular the authors
of \cite{BDF} have carried out the one-loop renormalization in a
generic $\xi$-gauge: on-shell the dependence on $\xi$ cancels
together with all unphysical thresholds, but off-shell there is a
host of unphysical thresholds depending on $\xi$. In view of the
extension of the SD approach to non-commutative gauge theories,
explicit gauge-independence of the off-shell propagator is
certainly worth to be obtained. We will show that the pinch
techniques can be extended to the non-commutative case, and give a solid check
of unitarity for the off-shell propagators by computing the
diagram that are  connected to it by the optical theorem.
Concerning the presence of the tachyonic pole and its cure by
softly-breaking ${\mathcal N}=4$ SUSY, we confirm the analysis
presented in \cite{cpt} and  \cite{KV}  where the computation was carried out
within the background field method. This should not come as a
surprise because the pinched propagator coincides with the
background gauge field one for $\xi_Q=1$ in the commutative setup.

New results are obtained when discussing the two-dimensional case.
Gauge theories in two dimensions do not have propagating local
degrees of freedom and this property should survive the
non-commutative deformation: at the classical level, by simply
choosing any axial gauge, the non-commutative $U(1)$ theory
reduces to its commutative cousin, that is a trivial
non-interacting theory. On the other hand, it is known that the
relation between perturbative and non-perturbative aspects is
subtle for $2D$ gauge theories \cite{bacci,bacc} and
non-commutativity has already produced some surprises when
computing Wilson loops \cite{bacc1,mak,bacc2,rick}. In the present
case, we would expect naively no correction to the free-propagator
when matter is absent, even in the non-commutative case, due to
the gauge-invariant meaning of the pinched self-energy. We obtain
instead a surprising result: working in a covariant gauge, the
dimensionally regularized theory, in the limit $D\rightarrow 2 $,
exhibits, even in absence of matter,
 a non-trivial $\theta$-dependent correction
to the dispersion relation which owes its finiteness to a fine
cancellation between planar and non-planar contribution.
 Moreover as $\theta\to 0$, we observe an
anomalous behavior from the  matter contributions, apparently
inducing a mass for the photon: this is analogous to what happens
in three-dimensions for the Chern-Simons term generated by
Majorana fermions \cite{chu}. On the other hand, when $\theta\to
\infty$ the original infrared divergences, tamed by
non-commutativity, reappear, leaving us with a ultraviolet
logarithmic divergent term. This is a twisted incarnation of the
UV-IR effect. We remark that our result should be meaningful, due
to the use of the gauge-invariant pinched self-energy.

The plan of the paper is the following: in section 2 we describe
how to extend the pinch techniques to the non-commutative case
and we present the computation of the
``pinched" gluon self-energy. 
In section 3 we  check the on-shell behavior of the pinched
propagator, we study  unitarity in the spacelike case by  proving  the optical theorem
 and discuss   analyticity by analyzing
the dispersion
 relation\footnote{
  We warn the reader about the two meanings of the term ``dispersion relation''
  which are used here and in the literature.
  On the one hand, it can be taken to mean the dependency of a particle's
  energy on its momentum, $E^2(\vec p)$, which becomes non-trivial because
  of the breaking of Lorentz invariance.
  The second use refers to the relation between the real and imaginary
  parts of an analytic function. The distinction is always clear from the context.
} connecting the real and imaginary parts of the gluon
self-energy. Section 4 is devoted to  study  the
two-dimensional gauge theory, with and without matter.
 Section 5 contains some concluding
 remarks and appendices are devoted to some technical aspects of the
computations.

\section{Pinch techniques}

One of the main problems in quantizing  gauge theories
 is to deal with the unphysical poles and
thresholds that generically plague local Green functions. On a
theoretical ground this is not an issue:  the well-known answer is
to limit the analysis to gauge-invariant quantities, which are
free of such troubles. On a practical ground, however, the
question  remains. In fact, simple invariant quantities like the
$S$-matrix elements are only defined on-shell, and the off-shell
physics is out of their grasp. Besides, more general off-shell
observables like the Wilson and Polyakov loops are non-local and
this makes them much more difficult to compute.

Pinch techniques (PT), in the original formulation by Cornwall
\cite{lots,Cornwall}, provide a manageable solution to this
problems. These consist in an algorithm that rearranges the
$S$-matrix elements of gauge theories and produces off-shell
proper correlation functions which satisfy the same Ward
identities (WI) as those produced by the classical Lagrangian
\cite{Cornwall,PT1}. The PT off-shell Green functions, in addition
to being gauge invariant by construction, also satisfy basic
theoretical requirements such as unitarity, analiticity and
renormalizability. They can be also used as the building blocks
of gauge-invariant Schwinger--Dyson equations, which allow to
discuss non perturbative questions as vacuum stability, dynamical
mass generation \cite{Cornwall}, and the behavior of unstable
states \cite{PT2} in the commutative setup. A comprehensive
discussion of this topic is out of the goals of the present paper
(see  \cite{review} for excellent reviews); here, instead, we
shall briefly outline how these techniques have been applied in
commutative theories to the case we are interested in: the
one-loop vacuum polarization.

In a nutshell, PT  consists in a judicious use of the
cancellations that underlie the well-known gauge invariance of the
$S$-matrix. One can concentrate on a two-particle process, and
identify which scattering amplitude contains the relevant
information about self-energy by simply looking at the structure
of the exchanged momenta. The classical choice is the
fermion-antifermion scattering process $f (p_1) + \bar f(p_2)\to
f^\prime(k_1)+ \bar{f}^\prime(k_2)$, but any other two-particle
process would be adequate, since the result does not depend on
this choice \cite{fourscat}. It is convenient to paramterize the
one-loop amplitude $f\bar f\to  f^\prime \bar{f}^\prime$ in any
covariant $\xi$-gauge\footnote{With
  this expression, we mean the usual gauge breaking term given by $\frac{1}{2\xi}(\partial_\mu A^\mu)^2$.}
so that the structure of the exchanged momenta is most evident:
\begin{eqnarray}
\label{pinch1}
    \langle f(p_1) \bar f(p_2)|T(s,t) |f^\prime(k_1)\bar f^\prime(k_2)\rangle
    &=&
     \Gamma^\mu(p_1,p_2) \Delta^{(\xi)}_{\mu\alpha}(s)
     \Pi^{(\xi)\alpha\beta}(s)
     \Delta^{(\xi)}_{\beta\nu}(s) \Gamma^\nu(k_1,k_2)
    +\nonumber\\&+&
    \Gamma_1^{(\xi)\mu}(p_1,p_2) \Delta^{(\xi)}_{\mu\nu}(s) \Gamma^\nu(k_1,k_2)
    +\\&+&
    \Gamma^\mu(p_1,p_2) \Delta^{(\xi)}_{\mu\nu}(s) \Gamma_1^{(\xi)\nu}(k_1,k_2)
    +{B}^{(\xi)}(p_1,p_2,-k_1,-k_2),\nonumber
\end{eqnarray}
where $s$ and $t$ are Mandelstam variables. In (\ref{pinch1}) the
symbol $\Pi^{(\xi)\alpha\beta}(s)$ designates the gauge-dependent
vacuum polarization, and $ \Gamma_1^{(\xi)\nu}$ and ${B}^{(\xi)}$
denote respectively the one-loop correction to the cubic $\bar{f}
A {f}$ vertex and box graphs; $\Delta^{(\xi)}_{\mu\nu}(s)$ and
$\Gamma^\mu$ stand for the  tree-level gluon propagator  and
vertex. External current are included in vertices
$\Gamma^{(\xi)\nu}$ and $\Gamma_1^{(\xi)\nu}$. Finally, the
superscript $(\xi)$ over each term  in (\ref{pinch1}) marks the
intrinsic gauge dependence of the different contributions. The
above decomposition corresponds to summing the diagrams depicted
in fig. \ref{cip2323}. \FIGURE{ \hspace{-3.3cm} { \SetScale{0.7}
  \SetScaledOffset(0,0)
\begin{picture}(80,80)(-60,-40)
    \ArrowLine(-60,-35)(-50,0)
    \ArrowLine(-50,0)(-60,35)
    \Gluon(-50,0)(-15,0){5.0}{3}
    \COval(0,0)(15,15)(0){Black}{Gray}
    \Gluon(15,0)(50,0){5.0}{3}
    \ArrowLine(60,-35)(50,0)
    \ArrowLine(50,0)(60,35)
     \Text(53,0)[]{$+$}
     \Text(0,-42)[]{(a)}
\end{picture}}\hskip .5cm
{   \SetScale{0.7}
  \SetScaledOffset(0,0)
\begin{picture}(30,80)(-65,-40)
    \ArrowLine(-60,-35)(-50,0)
    \ArrowLine(-50,0)(-60,35)
    \COval(-50,0)(15,15)(0){Black}{Gray}
    \Gluon(-35,0)(10,0){5.0}{5}
    \ArrowLine(20,-35)(10,0)
    \ArrowLine(10,0)(20,35)
     \Text(25,0)[]{$+$}
     \Text(-17,-42)[]{(b)}
\end{picture}}
{   \SetScale{0.7}
  \SetScaledOffset(0,0)
\begin{picture}(60,80)(-65,-40)
    \ArrowLine(-20,-35)(-10,0)
    \ArrowLine(-10,0)(-20,35)
    \Gluon(-10,0)(35,0){5.0}{5}
    \ArrowLine(60,-35)(50,0)
    \ArrowLine(50,0)(60,35)
    \COval(50,0)(15,15)(0){Black}{Gray}
     \Text(60,0)[]{$+$}
     \Text(12,-42)[]{(\={b})}
\end{picture}}
\hskip 1.5cm {   \SetScale{0.7}
  \SetScaledOffset(0,0)
\begin{picture}(20,80)(-65,-40)
     \ArrowLine(-60,-40)(-60,0)
     \ArrowLine(-60,0)(-60,40)
     \ArrowLine(10,0)(10,40)
     \ArrowLine(10,-40)(10,0)
     \Gluon(-60,20)(10,20){5.0}{7}
     \Gluon(-60,-20)(10,-20){5.0}{7}
      \Text(18,0)[]{$+$}
      \Text(-22,-42)[]{(c)}
\end{picture}}
\hskip 1.2cm {   \SetScale{0.7}
  \SetScaledOffset(0,0)
\begin{picture}(20,80)(-65,-40)
\ArrowLine(-60,-40)(-60,0)
     \ArrowLine(10,0)(10,40)
     \ArrowLine(10,-40)(10,0)
     \ArrowLine(-60,0)(-60,40)
     \Gluon(-60,20)(10,-20){5.0}{7}
     \Gluon(-60,-20)(10,20){5.0}{7}
     \Text(-22,-42)[]{(\={c})}
\end{picture}}
\caption{\label{cip2323} The contribution $(a)$ corresponds to
$\Pi^{(\xi)\alpha\beta}$; $(b)$ and $(\bar{b})$ contain the
one-loop correction ${\Gamma}^{(\xi)\mu}$ to the vertex, and
$(c)$,$(\bar{c})$ are the box diagrams $B^{(\xi)}$.} }

Gauge-invariance of the $S$-matrix ensures that the  sum of all
the graphs $(a)+(b)+(\bar{b})+(c)+(\bar{c})$ of figure
\ref{cip2323} is independent of the gauge  parameter $\xi$, and so
the matrix element $\langle f(p_1) \bar f(p_2)|T(s,t)
|f^\prime(k_1)\bar f^\prime(k_2)\rangle$ must also be independent
of $\xi$. The cancellation occurring between these different
diagrams is well-known but rather intricate, and it has some
surprises in store. By examining the analytical structure of this
scattering amplitude one can easily identify a few different
sectors: some terms depend solely on $s$, and there are other,
more complicated terms that carry an intrinsic dependence both on
$s$ and $t$. This suggests that the cancellations responsible for
the invariance of (\ref{pinch1})  are not ``global'', but occur
separately in different channels, so that more than one invariant
structure is buried there.

That this naive observation leads to a concrete and useful
application is non-trivial. Cornwall \cite{Cornwall} has indeed
shown that it is possible to rearrange (\ref{pinch1}) in the
following form
\begin{eqnarray}
\label{pinch2}
    \langle f(p_1)\bar f(p_2)|T(s,t) |f(k_1)\bar f(k_2)\rangle
    &=&
     \Gamma^\mu(p_1,p_2) \Delta^{(\xi)}_{\mu\alpha}(s)
     \widehat\Pi^{\alpha\beta}(s)
     \Delta^{(\xi)}_{\beta\nu}(s) \Gamma^\nu(p_3,p_4)
    +\nonumber\\
    &+&\widehat{\Gamma}^{\mu}(p_1,p_2) \Delta^{(\xi)}_{\mu\nu}(s) \Gamma^\nu(k_1,k_2)
    +\\&+&
    \Gamma^\mu(p_1,p_2) \Delta^{(\xi)}_{\mu\nu}(s) \widehat{\Gamma}^{\nu}(k_1,k_2)
    +\widehat{B}(p_1,p_2,-k_1,-k_2) ,\nonumber
\end{eqnarray}
where the \textit{pinched} polarization tensor
$\widehat{\Pi}^g_{\alpha\beta}(s)$, vertex $\widehat{\Gamma}^\mu$
and box $\widehat{B}$ are separately gauge-invariant and
independent of $\xi$. The surviving dependence on $\xi$ in
$\Delta^{(\xi)}_{\mu\nu}(s)$ is irrelevant, and it drops out as
soon as the tree level propagator hits the external current in
$\Gamma^\mu$ or $\hat \Gamma^\mu$; it has been left just for
future convenience.

Practically, the pinched representation (\ref{pinch2}) is obtained
in two steps. First one extracts from the second and third lines
of (\ref{pinch1}) those contributions which depend solely on the
Mandelstam variable $s$. Then one combines these contributions
with $\Pi^{(\xi)\alpha\beta}$ to yield a gauge-invariant quantity
$\widehat{\Pi}^g_{\mu\nu}$. Clearly, some more work is required in
the second and the third line to single out $\widehat{\Gamma}$ and
$\widehat{B}$, but as long as one is interested only in the pinched
vacuum polarization, this step is not relevant. More details can
be found in the literature.

The experienced reader may doubt that the procedure we just
sketched out is uniquely determined. In fact, since we are dealing
with scattering amplitudes, any redefinition that is  proportional
to the equations of motion leaves equation (\ref{pinch2})
unaltered. To fix this ambiguity and consistently promote
off-shell the pinched Green functions one imposes, apart from a
$\xi$-independence, a few reasonable constraints:
\begin{enumerate}
  \item The resulting Green's functions must be free from unphysical
  poles and thresholds.
  \item The Green functions must satisfy the
  tree-level Ward identities dictated by the classical Lagrangian.
  \item
  The Green functions must be resummable and compatible with the off-shell
  Schwinger--Dyson equations. This means, for example, that the
  $\xi$-dependence must cancel before integrating over
  loop momenta.
  \item The resummed Green functions must revert,
  when evaluated on-shell, to the conventional ones. This means that the
  resummation prescription must leave the position of the
  poles unchanged, since this is a gauge-invariant information.
\end{enumerate}
These constraints make the pinched Green's functions uniquely
defined.

In the commutative setup there would be additional items in this
list, making reference to the constraints dictated by unitarity
and analyticity. Since we deal with non-commutative field theories,
where unitarity might be jeopardized by nonlocal effects, we shall
drop these requirements and proceed without imposing them. As we
will show, the outcome of this approach is threefold: first of
all, it shows the applicability of the pinch techniques in the
non-commutative framework, even though the additional constraints
of unitarity and analyticity are not imposed. Secondly, it
provides us with a gauge-invariant test of unitarity for the
non-commutative theory; and finally, it allows for an investigation
of the analyticity properties through the analysis of the
dispersion relation.

We are now ready to illustrate some details of the computations
leading to the pinched non-commutative vacuum polarization. As is
well-known in the literature, the tensorial structures involved in
the one-loop \textit{un}integrated amplitudes are unchanged by
non-commutativity, even though Lorentz invariance is
broken\footnote{A new tensorial structure will emerge only after
integrating over loop momentum.}. The only difference consists in
the presence of a trigonometric factor inside the verteices that
spreads the nonlocal information, and depends on the loop and
external momenta (see Appendix A for the Feynman rules). This new
ingredient, however, does not impair the pinching procedure, and
it can be examined separately as we now show. Consider, for
example, the first diagram in fig. \ref{Smatrixpinchfigure}.
Employing the Feynman rules of Appendix A for the gauge fields and
Dirac fermions, the trigonometric factors associated with this
amplitude are \bea
  \label{mona}
  e^{i p_1\theta p_2/2 -i (k_2-k_1)\theta r/2}\sin{\left(\frac{q\theta\ell_2}{2}\right)}
  &=&
  e^{i p_1\theta p_2/2 -i (k_2-k_1)\theta (k_2-\ell_2)/2} \sin{\left(\frac{q\theta\ell_2}{2}\right)}
  =\nonumber\\&=&
  e^{i p_1\theta p_2/2 +i k_1\theta k_2/2}\left[e^{-i q\theta \ell_2} \sin{\left(\frac{q\theta\ell_2}{2}\right)} \right].
\eea This contains an overall factor associated with the
scattering of two fermions, which depends only on the external
momenta, and another quantity, in brackets, which is relevant for
the loop integration. However when taking into account the
contribution of the ``mirror" diagram (the second one in figure
\ref{Smatrixpinchfigure}), one sees that there the dependence on
the phase factors is got by simply setting $q\to -q$ in
(\ref{mona}).
Summing the pinch contributions of this two diagrams, the
trigonometric factors recombine into \beq
  2 ie^{i p_1\theta p_2/2 +i k_1\theta k_2/2}\left[
  \sin^2{\left(\frac{q\theta\ell_2}{2}\right)}\right].
\eeq The quantity in brackets has the right structure to mimics
the phase factor of the  vacuum polarization, so it can be
recombined with it to yield a pinched polarization tensor. The
process we just outlined repeats itself unaltered for all the
other diagrams. Thus, in the following, we shall focus just on the
tensorial structure, and we shall reinsert the necessary
trigonometric factors upon integrating.

The next step consists in extracting the $s$-dependent
contributions from diagrams $(a)$, $(b)$ and $(c)$ as shown in
figure \ref{Smatrixpinchfigure}.
\FIGURE{
   {\SetScale{0.6}
  \SetScaledOffset(0,-60)
  \begin{picture}(130,80)(-65,-40)
    \ArrowLine(-60,-35)(-40,-20) \Text(-30,-64)[]{$k_1$}
    \ArrowLine(-40,-20)(-40,20)\Text(-30,-34)[]{$r$}
    \ArrowLine(-40,20)(-60,35)\Text(-30,-8)[]{$k_2$}
    \Gluon(-40,-20)(-10,0){5.0}{5}\Text(-10,-19)[]{$\ell_2$}
    \Gluon(-40,20)(-10,0){5.0}{5}\Text(-10,-52)[]{$\ell_1$}
    \Vertex(-10,0){1}
    \Gluon(-10,0)(40,0){5.0}{5}\Text(7,-25)[]{$q$}
    \ArrowLine(60,-35)(40,0)\Text(30,-64)[]{$p_1$}
    \ArrowLine(40,0)(60,35)\Text(30,-8)[]{$p_2$}
  \end{picture}}
  $\!\!\!\!\!\rightarrow\  \ \ $
  \ScaledPinchedTriGraph{0.6}{0}{-60}{$\ \mathcal{I}_{pinch.}^{(a)}=$}
  \ScaledBoxGraph{0.6}{0}{-60}
  $\!\!\!\!\!\rightarrow\  \ \ $
  \ScaledPinchedBoxGraph{0.6}{0}{-60}{$\ \mathcal{I}_{pinch.}^{(b)}=$}
  \ScaledVertexGraph{0.6}{0}{-60}
  $\!\!\!\!\!\rightarrow\  \ \ $
  \ScaledPinchedVertexGraph{0.6}{0}{-60}{$\ \mathcal{I}_{pinch.}^{(c)}=$}
  \vspace{.5cm}
 \caption{
 \label{Smatrixpinchfigure}
 One-loop propagator-like contributions to the pinch technique.}
  }
This is accomplished by means of a classical trick. Let us denote
respectively with $r$, $\ell_1$ and $\ell_2$  the momenta of
fermion and of the two gluons running in the loop in the first
diagram in fig. \ref{Smatrixpinchfigure}. In addition, call the
external momenta attached to this loop   $k_1$ and $k_2$.  Then
the following silly Ward identities hold \bea
\label{trivialWardfermions}
  \ssh \ell_1 &=& (\ssh r - \ssh k_1)=(\ssh r-m)- (\ssh k_1-m)
   = S^{-1}(r) - S^{-1}(k_1)\nonumber\\
   \ssh \ell_2  &=& (\ssh k_2 -\ssh r) =(\ssh k_2 -m)-(\ssh r-m)
   = S^{-1}(k_2) - S^{-1}(r),
\eea where $S$ is the free fermion propagator. Equations
(\ref{trivialWardfermions}) state that any term of the form $\ssh
\ell_1 $ or $\ssh \ell_2 $ present in the diagrams spawns two
terms: one is proportional to the equation of motion of the
external leg, and it can be dropped; the second one is
proportional to the inverse of the propagators of the fermion
running inside the loop. The net effect of this procedure is to
squeeze away ``pinch'' the internal fermionic propagators.
Graphically, this mechanism is represented by the diagrams
appearing on the r.h.s. of fig. \ref{Smatrixpinchfigure}. 
We have obtained  effective Feynman diagrams where one or both of the
fermion propagators have been pinched: these diagrams exhibit
clearly their dependence on merely $s$. The same trick allows us
to handle the other two diagrams in fig. \ref{Smatrixpinchfigure}.
In the present discussion we have neglected  diagrams governing
the renormalization of the external legs, since their total effect
on the pinching procedure vanishes.

Now we collect the different contributions to the pinched vacuum
polarization. We start from those coming from the diagram $(a)$ in
figure \ref{cip2323}. \FIGURE{\hspace{-1cm}
  \ScaledGlueLoop{0.9}{0}{0}{$\mathcal{I}_{gluon}$}
  \ScaledTadpole{0.9}{0}{0}{$\mathcal{I}_{tadpole}$}
  \ScaledGhostLoop{0.9}{0}{0}{$\mathcal{I}_{ghost}$}
  \vspace{.5cm}
  \caption{\label{ordinarycontributionstoPi} Ordinary contributions to $\Pi^{(\xi)}_{\alpha\beta}$.}
  }
These include the three ordinary contributions to the vacuum
polarization tensor of non-commutative QED, represented in figure
\ref{ordinarycontributionstoPi}. Their value is reported below:
\begin{eqnarray}
  \mathcal{I}_{gluon}&=&
  \left[
  \frac{g_{\alpha\beta}}{k^2}
  +\frac{2\,q^2\,g_{\alpha\beta}
  +(2D-3)k_{\alpha}k_{\beta}
  }{k^2p^2}
  +(1-\xi)
  \left(
     \frac{-k^2 g_{\alpha\beta}+ k_\alpha k_\beta}{p^4}
     -2\frac{q^2 k_\alpha k_\beta}{k^2p^4}
    +\right.\right.\nonumber\\&&\left.\left.
    +2\frac{q^2g_{\alpha\beta}}{k^4}
    -\frac{q^4 g_{\alpha\beta}}{k^2p^4}
  \right)
  +{(1-\xi) }^2
  \left(
    \frac{q^4 k_\alpha k_\beta}{k^4p^4}
  \right)
  \right]
  +(k\leftrightarrow p),
\end{eqnarray}
\begin{eqnarray}
  \mathcal{I}_{ghost}=
  -\left(\frac{k_\alpha p_\beta+k_\beta  p_\alpha }{k^2 p^2}\right)
  \ \ \ \ \
  \mathcal{I}_{tadpole} =
  2(1-D)\frac{g_{\alpha\beta}}{k^2}
  -2(1-\xi)\left(\frac{g_{\alpha\beta}}{k^2}-\frac{k_\alpha
  k_\beta}{k^4}\right).
\end{eqnarray}
We have denoted with $q$ the external momentum and with $k,\,p$
the loop momenta. Then we have pinched contributions from the
$s$-parts of diagrams  $(b)$ and $(c)$, as shown in figure
\ref{Smatrixpinchfigure}.
\begin{eqnarray}
  \mathcal{I}_{pinch.}^{(a)}&=&
   4q^2\frac{g_{\alpha\beta}}{k^2\,p^2}
   +2(1-\xi)
   \left[
    \frac{q^2\,k_\beta\,k_\alpha}{k^4\,p^2}
    +\frac{q^2\,p_\alpha\,p_\beta}{k^2\,p^4}
    +\left(\frac{q^4}{k^2\,p^4}+\frac{q^4}{k^4\,p^2}\right)g_{\alpha\beta}-\right.
    \nonumber\\ && \left.
    -\left(\frac{1}{k^4}+\frac{1}{p^4}\right)q^2g_{\alpha\beta}
   \right]
   -2(1-\xi)^2\left(
   \frac{q^4\,p_\alpha\,p_\beta}{k^4\,p^4}\right),\\
  \mathcal{I}_{pinch.}^{(b)}&=&
  2(1-\xi)  \left(
    - \frac{q^4 g_{\alpha\beta} }{k^2 p^4}
    - \frac{q^4 g_{\alpha\beta} }{k^4 p^2}
  \right)
  +2 {(1-\xi) }^2\frac{q^4 p_\alpha p_{\beta}}{k^4 p^4},\\
  \mathcal{I}_{pinch.}^{(c)}&=&
  -2(1-\xi)\left(
   \frac{q^2 g_{\alpha\beta}}{k^4}
  +\frac{q^2 g_{\alpha\beta}}{p^4}
  \right).
\end{eqnarray}
We remark that these are all independent of the dimension $D$.
Summing up all these contributions we find that the the pinched
polariszation tensor is
\begin{eqnarray}
\label{unintegratedpinchedPi}
    \!\!\!\!\!\!\!\!\!\!\!\!\widehat\Pi_{\alpha\beta}
    &=&
     -2 g^2
\int \frac{\dd^D k}{(4\pi)^{D/2}}
     \frac{8  (q^2g_{\alpha\beta} -q_\alpha q_\beta)+(4-2D)  (k^2g_{\alpha\beta} -
     k_\alpha k_\beta)      }{k^2(k+q)^2}\sin^2\left(\frac{q\theta k}{2}\right),
\end{eqnarray}
where we have restored the relevant trigonometric factor. Equation
(\ref{unintegratedpinchedPi}) is particularly intriguing not only
because it is manifestly transverse, but also because it signals
the possible presence of \textit{evanescent terms} in two
dimensions: there is a potential competition between the second
integral, which is logarithmically ultraviolet divergent, and its
vanishing coefficient when approaching $D=2$. This issue will be
discussed in detail in Sect.4.

The computations leading to the Euclidean version of the integrals
in (\ref{unintegratedpinchedPi}) are shown in appendix B. The
final result takes on the form
\begin{eqnarray}
\label{integratedpinchedPianyd}
 \widehat\Pi^g_{\mu\nu}=
  \widehat\Pi_c(q^2,|\tilde{q }|^2)\!\!
  \left(
    g_{\mu\nu}-\frac{q_\mu q_\nu}{q^2}
    -\frac{\tilde{q }_\mu \tilde{q }_\nu}{|\tilde{q }|^2}
  \right)\!\!
   +\!\!\left(
     \widehat\Pi_\theta(q^2,|\tilde{q }|^2)
    +\widehat\Pi_c(q^2,|\tilde{q }|^2)
  \right)
  \frac{\tilde{q }_\mu \tilde{q }_\nu}{|\tilde{q }|^2}.
\end{eqnarray}
Where $\tilde{q}^\nu=\theta^{\mu\nu} q_\nu$ and $ |\tilde{q}|^2=q
\bullet q = \Theta_E (p_0^2+p_1^2)+\Theta_B(p_2^2+p_3^2)$. The
notation here is chosen  to underline the existence of two
spacetime invariants for non-commutative gauge theories, $q^2$ and
$|\tilde{q }|^2$. In the following, to lighten the notation, the
dependence on the two invariants is understood. The explicit
values of the two functions are given by the following integrals
over the Feynman parameters
\begin{eqnarray}
    \widehat\Pi_{c}
    &=&
      -\frac{g^2}{(4\pi)^{D/2}}\!\!
    \int_0^1\!\! \dd x
    \frac{8q^2+(4-2D)(-M^2+q^2 x^2)}{(M^2)^{2-D/2}}
    \!\!\left(
    \Gamma(2-\frac{D}{2})-2\left(\frac{|\tilde{q}| M}{2}\right)^{2-\frac{D}{2}}\!\!\!\!\!\!\!\!\!\!\!
    K_{2-\frac{D}{2}}(M |\tilde{q}|)
    \right),\nonumber\\
    \label{pinchedPithetaanyd}
    \widehat\Pi_{\theta}
    &=&
      g^2\frac{(4-2D)}{(4\pi)^{D/2}}
  2  \int_0^1 \dd x \;
    2M^2\left(\frac{|\tilde{q }|}{2M}\right)^{2-D/2}  K_{-D/2}(M |\tilde{q
    }|),
\end{eqnarray}
where $M\equiv\sqrt{ x(1-x)q^2}$ and $K_n$ is the modified Bessel function  of the second kind.

\section{The four-dimensional theory}

We remarked above that in the commutative case one further
requires, among the defining properties of pinch-technique
resummed amplitudes, the off-shell optical relations, analyticity
and invariance of the position of the poles. In the
non-commutative case the situation is more involved because such
properties could be spoiled by non-commutative effects. This
section is devoted to a detailed analysis of these issues. We
begin by studying the on-shell properties of the pinched
propagator in the supersymmetric extension of the theory: as we
discussed above it is an important consistency check that the
on-shell physics is untouched by the pinching procedure. In the
following sections we discuss the unitarity and analyticity
properties of the pinched propagator, reverting to the matterless
case.

\subsection{Dispersion relations in the supersymmetric extension of the theory}

It is well known that in four dimensions the UV/IR mixing induces
a quadratic IR divergence in the self-energy, which produces a
tachyonic divergence in the dispersion relation. As observed in
\cite{cpt},\cite{KV}, one can recover vacuum stability by
introducing a sufficient amount of supersymmetric matter: SUSY in
fact, by improving the UV behavior of quantum loops, acts as a
regulator for the infrared divergences of a non-commutative
theory, because the UV effects are tamed before they can couple to
$\theta_{\mu\nu}$ and induce large-distance divergences. Adding
the contribution of $n_f $ fermions and $n_s$ scalars we find
\begin{eqnarray}
\label{susypi}
    \widehat\Pi_{c}
    &=&
    \frac{g^2}{(4\pi)^{D/2}}
    \int_0^1 \dd x \;\Bigg[
    -\frac{[2(2-D)(-M_g^2+q^2 x^2)+8q^2]}{(M_g^2)^{2-D/2}}A(m_g)+\nonumber\\
    &+& \frac{2\,\, D\,\, q^2 x (1-x)}{(M_f^2)^{2-D/2}}\sum_{f}A(m_f)
    +\frac{ q^2 (4 x^2-1)}{(M_s^2)^{2-D/2}}\sum_sA(m_s)\Bigg],
\\
    \widehat\Pi_{\theta}
    &=&
    \frac{g^2}{(4\pi)^{D/2}}
    \int_0^1 \dd x \;
    \Bigg[2(4-2D)B(m_g)
    +2D \sum_{f}B(m_f)
    -4\sum_s B(m_s)\Bigg]\label{susypi1}.
\end{eqnarray}
Here we have included softly supersymmetry-breaking masses $m_f$,
$m_s$, $m_g$ for fermions, scalars and gluons. The dependence on
the masses is contained in the functions $M_i$:
\begin{eqnarray}
    A(m_i)&=&
    \left[
    \Gamma(2-\frac{D}{2})-2\left(\frac{|\tilde{q}| M_i}{2}\right)^{2-\frac{D}{2}}\!\!\!\!\!\!\!\!\!\!\!
    K_{2-\frac{D}{2}}(M_i |\tilde{q}|)
    \right],
    \\
    B(m_i)&=&
     2M_i^2\left(\frac{|\tilde{q }|}{2M_i}\right)^{2-D/2}  K_{-D/2}(M_i |\tilde{q }|),
\end{eqnarray}
where $M_i(m_i,q)=m_i^2+x(1-x)q^2$. The planar part of the vacuum
polarization is UV-divergent for $D\rightarrow 4$ and needs to be
renormalized. We will perform this procedure just like in the
commutative case: we choose a subtraction scale $\mu$, so that
\begin{eqnarray}
A(m_i)&\rightarrow&
    -\left[
    \log
    \left(\frac{M_i}{\mu}\right)+K_0(M_i |\tilde{q}|)
    \right].
\end{eqnarray}
Two limits of these functions are worthy of being considered.

One of the virtues of the pinch techniques is their ability to
compute the $\beta$-functions, and we would like to recover these
in the appropriate limit. This limit consists in reaching the
ultraviolet by taking the arguments of the Bessel functions to be
large, so that their contribution is exponentially suppressed,
together with $q\gg m_i$:
\begin{eqnarray}
    \widehat\Pi_{c}
    =
  \frac{1}{4\pi^2}
    \left(\frac{11}{3} -\frac{2}{3}n_f -\frac{1}{6} n_s\right)
    \log\left(\frac{q}{\mu}\right).
\end{eqnarray}
We recover the standard $\beta$-function for softly broken susy gauge
theories: it is a further check of the validity of the PT
resummation prescription.

Of course one could compute the $\beta$-function using the
\textit{background field method} (BFM); however, as pointed out in
\cite{PT2}, BFM $n$-point functions display a residual dependence
on the gauge parameter $\xi_Q$ employed in fixing the gauge for
the quantum fields inside the loops, and this may lead to
unphysical thresholds. Requiring the absence of such unphysical
effects forces the choice $\xi_Q=1$, which cannot be otherwise
motivated; in this case the one-loop $n$-point functions evaluated
in BFM and the ones we computed using the pinch techniques
coincide.

A second limit concerns approaching the infrared with small
arguments of the Bessels, so that:
\begin{eqnarray}
A(m_i)&\rightarrow&
    \log\left(|\tilde{q}| \mu\right).
\end{eqnarray}
In this limit we have:
\begin{eqnarray}
    \widehat\Pi_{c}
    &=&
    -\frac{1}{4\pi^2}
    \left(\frac{11}{3}-\frac{2}{3}n_f -\frac{1}{6} n_s\right)
    \log \left(|\tilde{q}|\mu\right).
\end{eqnarray}
As already pointed out in \cite{cpt}, this expression shows that
the running of the coupling constant in the infrared is similar to
the one in the ultraviolet. A different sign indicates that the
theory becomes weakly coupled at low energy. The duality $q
\rightarrow \frac{1}{|\tilde{q}|}$ is thus interpreted as another
incarnation of the UV-IR mixing. The expression we found for the
self-energy, through equations (\ref{susypi}) and (\ref{susypi1}),
coincides with the one found in \cite{cpt} and \cite{KV}. In
particular, the pure gluon contribution to the equation for the
position of the poles gives rise to the well-known tachyonic
dispersion relation. This last feature implies that the
PT-resummed amplitudes reduce to the unpinched value when
evaluated on-shell as they should. In other words, the resummation
prescription does not modify the position of the poles.

\subsection{Optical theorem and unitarity }

Having verified the on-shell properties of the pinched propagator
we move to the analysis of the off-shell physics. Let us first of
all show how the pinch techniques can be employed to check the
optical theorem for \textit{off-shell} two-point functions. For
on-shell matrix elements the optical theorem states that if the
$S$-matrix is unitary then:
\begin{eqnarray}
  \ima \langle q \bar q
  |T|
  q \bar q \rangle
  =
  \frac{1}{2}\left(\frac{1}{2}\right)
  \int d\Omega \langle q \bar q
   |T|
   g  g \rangle \langle g  g
   |T|
   q \bar q \rangle.
\end{eqnarray}
On the left hand side we have the $S$-matrix element for a
$q$-$\bar q$  scattering process (the one from which the pinched
vacuum polarization is obtained), and on the right hand side we
have the amplitudes for quark-gluon scattering. The
diagrams contributing to this amplitude are obtained  by cutting
through the $S$-matrix and are  displayed in figure
\ref{opticalstuff}. The factor of $1/2$ appears because the final
on-shell gluons are identical particles.

\FIGURE{
  \begin{picture}(170,100)(-30,0)
    \Text(25,90)[r]{$p_{1}\rightarrow$}
    \Text(25,10)[r]{$p_{2}\leftarrow$}
    \ArrowLine(30,90)(50,50)\ArrowLine(50,50)(30,10)
    \Gluon(50,50)(90,50){3}{4}
    \CCirc(90,50){1}{Black}{Black}
    \Gluon(90,50)(110,90){3}{4}\Gluon(90,50)(110,10){3}{4}
    \Text(115,10)[l]{$\leftarrow k_{2},\beta$}
    \Text(115,90)[l]{$\leftarrow k_{1},\alpha$}
  \end{picture}
  \begin{picture}(170,100)(-30,0)
    \Text(25,90)[r]{$p_{1}\rightarrow$}
    \Text(25,10)[r]{$p_{2}\leftarrow$}
    \ArrowLine(30,90)(70,70)\ArrowLine(70,70)(70,30)\ArrowLine(70,30)(30,10)
    \Gluon(70,70)(110,90){3}{4}\Gluon(70,30)(110,10){3}{4}
    \Text(115,10)[l]{$\leftarrow k_{2},\beta$}
    \Text(115,90)[l]{$\leftarrow k_{1},\alpha$}
  \end{picture}
  \begin{picture}(170,100)(-30,0)
    \Text(25,90)[r]{$p_{1}\rightarrow$}
    \Text(25,10)[r]{$p_{2}\leftarrow$}
    \ArrowLine(30,90)(70,70)\ArrowLine(70,70)(70,30)\ArrowLine(70,30)(30,10)
    \Gluon(70,70)(110,10){3}{6}
    \CCirc(84,50){9}{White}{White}
    \Gluon(70,30)(110,90){3}{6}
    \Text(115,10)[l]{$\leftarrow k_{2},\beta$}
    \Text(115,90)[l]{$\leftarrow k_{1},\alpha$}
  \end{picture}
  \caption{ $s$- $t$- $u$-channel amplitudes}
  \label{opticalstuff}
}

In the last section we built a gauge-independent self-energy by
resumming all the one-loop $s$-channel contributions to the matrix
elements on the left hand side. The analogue of this procedure on
the right-hand side consists in recasting it as a sum of
gauge-independent $s$, $t$ and $u$-channel contributions. If the
two sides match, we will obtain a direct check of unitarity in the
$s$-channel. Let then ${\cal M}$ be the $q\bar q$ scattering
$S$-matrix element, and ${\cal T}$ be the $q$-$g$
scattering amplitude:
\begin{eqnarray}
 {\cal M}\doteq \langle q \bar q      |T|         q \bar q \rangle
 &,
 &
 {\cal T}\doteq \langle q \bar q      |T|         g  g \rangle.
\end{eqnarray}
${\cal T}$ consists of the ${\cal T}_s$, ${\cal T}_t$ and ${\cal
T}_u$ contributions displayed in figure \ref{opticalstuff}: we
must take the squared modulus and sum over all physical gluon
polarizations.

The matching of different channels from ${\cal M}$ and ${\cal T}$
is non-trivial, we shall show, in fact, that the ``pinched optical
theorem" relates ${\cal M}_s$ to the whole of ${\cal T}_s$ plus
pieces from ${\cal T}_t$ and ${\cal T}_u$. On second thoughts this
comes as no surprise: by cutting through the box diagram one
obtains ${\cal T}_t$ and ${\cal T}_u$, and it is just to be
expected that if the box gives pinch contributions, so must ${\cal
T}_t$ and ${\cal T}_u$. In the following we specialize to  $D=4$
and adopt Minkowskian signature. In order to analytically continue
the results of the previous section one needs to send:
\begin{eqnarray*}
 p^2_E \rightarrow -p_M^2,\ \ \ \ &&\ \  \ \ \tilde{p}_E^2\rightarrow  -\tilde{p}_M^2,
 \\(p \bullet p)_E = \Theta_E
(p_0^2+p_1^2)+\Theta_B(p_2^2+p_3^2) &\rightarrow & (p \bullet p)_M
= \Theta_E (p_0^2-p_1^2)+\Theta_B(p_2^2+p_3^2).
\end{eqnarray*}
Our computation follows closely \cite{PT2}. The contributions are:
\begin{eqnarray}
  {\cal T}_s^{\alpha\beta}
  &=&
  J^\mu(p_1,p_2)
  \sin\frac{-p_2\theta p_1}{2}
  \Delta_{\mu\nu}(q)
  V^{\nu\alpha\beta}(q,k_1,k_2)
    \nonumber\\
  {\cal T}_{t+u}^{\alpha\beta}
  &=&
  J^{\alpha\beta}(p_1,p_1+k_1,p_2)
  \sin\fract{p_2\theta k_2}{2}
  \sin\fract{p_1\theta k_1}{2}
  +(k_1,\alpha\leftrightarrow k_2,\beta)
  ,
\end{eqnarray}
where $J^{\mu}(p_1,p_2)=g \bar{v}(p_2)\gamma^\mu u(p_1) $ and
$J^{\alpha\beta}(p_1,p_1+k_1,p_2)=g^2 \bar{v} (p_2)\gamma^\alpha
S_F(p_1+k_1) \gamma^\beta u(p_1)$, while
$V_{\nu\alpha\beta}(q,k_1,k_2)$ is the three-gluon vertex. The
optical theorem states that
\begin{eqnarray*}
  \ima \{
   {\cal M}_s
   +
   {\cal M}_t
   +
   {\cal M}_u
   \}
  =
  \int \frac{\dd  \Gamma}{4}
  \left({\cal T}_s^{\mu\nu}+{\cal T}_{t+u}^{\mu\nu}\right)
  P_{\mu\rho}(k_1) P_{\nu\sigma}(k_2)
  \left({\cal T}_s^*{}^{\rho\sigma}+{\cal T}_{t+u}^*{}^{\rho\sigma}\right)
  ,
\end{eqnarray*}
where ${\cal M}_s$ contains the pinched vacuum polarization tensor
$\widehat \Pi_{\mu\nu}$,
\begin{eqnarray}
  \ima {\cal M  }_{s}
  =
     \int d\Gamma  \frac{1}{4q^4}
  J^\mu(p_1,p_2)
  \widehat \Pi_{\mu\nu}
  J^{*\nu}(p_1,p_2)
   \sin^2 \frac{k_1\theta k_2}{2} \sin^2 \frac{p_1\theta p_2}{2}
\end{eqnarray}
and $P_{\mu\nu}$ represents the polarization tensor (the sum over
the gluons' physical polarizations)
\begin{eqnarray}
  P_{\mu\nu}(q,\eta)
  \doteq
    -g_{\mu\nu}
  +\frac{\eta_\mu q_\nu +q_\mu \eta_\nu }{(\eta\cdot q)}
  +\frac{\eta^2 q_\mu q_\nu}{(\eta\cdot q)^2}.
\end{eqnarray}

Before we plunge into the calculations, it is important to make a
few remarks concerning the consistence of the pinch techniques as
they are applied to the squared modulus of ${\cal T}$. One should
notice that in this case we have a dependence on two
``gauge-parameters": the gauge-fixing one, $\xi$, and $\eta_\mu$,
which is introduced upon choosing the two independent physical
polarization vectors to be summed over. We demand that the pinch
contributions from the second and third graphs cancel the $\xi$
and $\eta_\mu$-dependence of the first diagram. Let us begin by
showing the independence from $\xi$. By decomposing the
three-gluon vertex as $V^{T}=V^{F}+ V^{P}$ we have
\begin{eqnarray}
  V^{F}_{\mu\nu\rho}(q,p,k)&=& g\left[ (p-k)_\mu g_{\nu\rho} +2q_\nu
  g_{\mu\rho} -2q_\rho g_{\mu\nu} \right] \sin\frac{p\theta k}{2},
  \\
  V^{P}_{\mu\nu\rho}(q,p,k)&=&g
  [k_\rho g_{\mu\nu}-p_\nu g_{\mu\rho}]\sin\frac{p\theta k}{2}.
\end{eqnarray}
We observe that for on-shell gluons, which obey $k^\mu
P_{\mu\nu}=0$ and $k^2=0$, the term $V^P_{\mu\nu\rho}$ dies upon
hitting  $P_{\mu \rho}$  $P_{\nu \sigma}$. The $s$-channel's
explicit dependence on the $\xi$-gauge disappears thanks to the
fact that the gauge-fixing term is longitudinal, so it also drops
off as it hits the conserved external fermionic current\footnote{
 Had we not used a covariant gauge, current conservation
 would not have been sufficient to guarantee the gauge fixing
 independence.
 Like in the commutative case \cite{PT2},
 in this case one resorts to the following Ward identity
\[
  q^\mu V^F_{\mu\nu\rho}(q,-p-k)=(p^2-k^2)g_{\nu \rho}
  .
\]
}. Then the $s$-channel propagator boils down to its Feynman-gauge
values  and all dependence on $\xi$ vanishes. We must still prove
that also the dependence on $\eta_\mu$ gets cancelled as well.

Luckily, all the phases factorize exactly like in the first
section and the following set of Ward identities can be obtained:
 \begin{eqnarray*}
  k_1{}^\alpha
  ({\cal T}_s)_{\alpha\beta}
  &=&
  \left(
  2\frac{{k_1}^\mu {k_2}^\beta }{q^2}
  -g^{\mu\beta}
  \right)
  J^\mu(p_1,p_2)
  \sin\fract{p_1\theta p_2}{2}
  \sin\fract{k_1\theta k_2}{2},
\\
  k_2{}^\beta
  ({\cal T}_s)_{\alpha\beta}
  &=&
  \left(
  2\frac{{k_2}^\mu {k_1}^\alpha}{q^2}
  +g^{\mu\alpha}
  \right)
  J^\mu(p_1,p_2)
  \sin\fract{p_1\theta p_2}{2}
  \sin\fract{k_1\theta k_2}{2},
\\
  k_1{}^\alpha
  ({\cal T}_{t+u})_{\alpha\beta}
  &=&
  J_\beta(p_1,p_2)
  \sin\fract{p_1\theta p_2}{2}
  \sin\fract{k_1\theta k_2}{2},
\\
  k_2{}^\beta
  ({\cal T}_{t+u})_{\alpha\beta}
  &=&
  -J_\alpha(p_1,p_2)
  \sin\fract{p_1\theta p_2}{2}
  \sin\fract{k_1\theta k_2}{2},
\\
  k_1{}^\alpha
  k_2{}^\beta
  ({\cal T}_s)_{\alpha\beta}
  &=&
  -{k_2}_\mu J^\mu(p_1,p_2)
  \sin\fract{p_1\theta p_2}{2}
  \sin\fract{k_1\theta k_2}{2},
\\
  k_1{}^\alpha
  k_2{}^\beta
  ({\cal T}_{t+u})_{\alpha\beta}
  &=&
  k_2{}^\beta
  J_\beta(p_1,p_2)
  \sin\fract{p_1\theta p_2}{2}
  \sin\fract{k_1\theta k_2}{2}.
\end{eqnarray*}
Defining:
\begin{eqnarray}
  {\cal G}
  \doteq
    -\frac{k_2^\mu}{q^2}J_\mu(p_{\rm in},p_{\rm out})
  \sin\fract{p_1\theta p_2}{2}
  \sin\fract{k_1\theta k_2}{2}=\frac{k_1^\mu}{q^2}J_\mu(p_{\rm in},p_{\rm out})
  \sin\fract{p_1\theta p_2}{2}
  \sin\fract{k_1\theta k_2}{2},
\end{eqnarray}
we arrive at the relevant cancellation laws  displayed in
figure \ref{cancellationz}
\begin{eqnarray}
  k_1{}^\alpha
  ({\cal T}_{\rm tot})_{\alpha\beta}
  &=&
  2{k_2}^\beta {\cal G},
\\
  k_2{}^\beta
  ({\cal T}_{\rm tot})_{\alpha\beta}
  &=&
  2{k_1}^\alpha {\cal G},
\\
  k_1{}^\alpha
  k_2{}^\beta
  ({\cal T}_{\rm tot})_{\alpha\beta}
  &=&
  0.
\end{eqnarray}

\FIGURE{ \hspace{-40pt}

  \SetScale{0.5}
  \begin{picture}(65,50)(+5,20)
    \ArrowLine(30,90)(50,50)\ArrowLine(50,50)(30,10)
    \Gluon(50,50)(90,50){3}{4}
    \CCirc(90,50){1}{Black}{Black}
    \Gluon(90,50)(110,90){3}{4}\Gluon(90,50)(110,10){3}{4}

  \end{picture}
  $k_1^\mu
  =$
  \SetScale{0.5}
  \begin{picture}(65,50)(+5,20)
    \ArrowLine(30,90)(50,50)\ArrowLine(50,50)(30,10)
    \Gluon(50,50)(90,50){3}{4}
    \CCirc(90,50){1}{Black}{Black}
    \DashArrowLine(90,50)(110,90){4}
    \DashArrowLine(110,10)(90,50){4}
     \end{picture}
     $k_2^\nu
  +$
  \begin{picture}(65,50)(+5,20)
    \ArrowLine(30,90)(50,50)\ArrowLine(50,50)(30,10)
    \CCirc(50,50){4}{Black}{Black}
    \Gluon(50,50)(110,90){3}{4}\Gluon(50,50)(110,10){3}{4}
  \end{picture}
\\
  \SetScale{0.5}
  \Bigg(
  \begin{picture}(65,50)(+5,20)
    \ArrowLine(30,90)(70,70)\ArrowLine(70,70)(70,30)\ArrowLine(70,30)(30,10)
    \Gluon(70,70)(110,90){3}{4}\Gluon(70,30)(110,10){3}{4}
  \end{picture}
  $+$
  \begin{picture}(65,50)(+5,20)
    \ArrowLine(30,90)(70,70)\ArrowLine(70,70)(70,30)\ArrowLine(70,30)(30,10)
    \Gluon(70,70)(110,10){3}{6}
    \CCirc(84,50){9}{White}{White}
    \Gluon(70,30)(110,90){3}{6}
  \end{picture}
  $\Bigg)k_1^\mu
  =$
  \SetScale{0.5}
  $-$
  \begin{picture}(65,50)(+5,20)
    \ArrowLine(30,90)(50,50)\ArrowLine(50,50)(30,10)
    \CCirc(50,50){4}{Black}{Black}
    \Gluon(50,50)(110,90){3}{4}\Gluon(50,50)(110,10){3}{4}
  \end{picture}
  \caption{$s$- channel (first line), and $t$- and $u$- channels cancellations \label{cancellationz}}
  }
Thanks to this cancellation it is straightforward to recast the
optical theorem as
\begin{eqnarray}
  \ima \{
  {\cal M}_s
  +
  {\cal M}_t
  +
  {\cal M}_u
  \}
   =
   && \frac{1}{4}\int d\Gamma
  \Biggl[
  \underbrace{(
  {\cal T}_s^{\mu\nu}{\cal T}_s^*{}_{\mu\nu}
  -8{\cal G}{\cal G}^*)
  }_{\rm propagator-like}
  +  \underbrace{
  \left({\cal T}_s^{\mu\nu}{\cal T}_{t+u}^*{}^{\mu\nu}
  +{\cal T}_{t+u}^{\mu\nu}{\cal T}_s^*{}_{\mu\nu}\right)
  }_{\rm vertex-like}+\nonumber\\
  &&+\underbrace{{\cal T}_{t+u}^{\mu\nu}{\cal T}_{t+u}^*{}_{\mu\nu}
  }_{\rm box-like}
  \Biggr],
\end{eqnarray}
where all the dependence on the gauge parameters $\xi$ and
$\eta_\mu$ has vanished.

We have shown that a gauge-independent decomposition is possible,
and we have identified the contribution which ought to be related
by the optical theorem to $\widehat \Pi^{\mu\nu}$. We just need to
compute the other side of the optical theorem's equation: for this
we just need to compute
\begin{eqnarray}
  \ima {\cal M  }_{s}
  &=&\int
     \frac{  d\Gamma }{4q^4}
  J^\mu(p_1,p_2)
  \left[{\cal T}_s^{\mu\nu}{\cal T}_s^*{}_{\mu\nu}
  -8{\cal G}{\cal G}^*
  \right]
  J^{*\nu}(p_1,p_2)
   \sin^2 \frac{k_1\theta k_2}{2} \sin^2 \frac{p_1\theta p_2}{2}
  =\nonumber\\&=&
  \int d\Gamma
  \frac{1}{2}J^\mu(p_1,p_2)
  \frac{1}{q^2}
  \left[
  4q^2(g_{\mu\nu} -\frac{q^\mu q^\nu}{q^2})
  +\frac{(k_1^\mu-k_2^\mu) (k_1^\nu-k_2^\nu)}{q^4}
  \right]\times\nonumber\\
  &\times&
  \frac{1}{q^2}
  J^{*\nu}(p_1,p_2)
  \sin^2 \frac{k_1\theta k_2}{2} \sin^2 \frac{p_1\theta p_2}{2}.
\end{eqnarray}
We introduce ($p=k_1$ and $k_2=q-k_1$)
\begin{eqnarray}
  \frac{1}{2}  A^{\mu\nu}
  \doteq
  4q^2\left( g^{\mu\nu}-\frac{q^\mu q^\nu}{q^2}\right)+(2p-q)^\mu (2p-q)^\nu,
\end{eqnarray}
in terms of which the optical theorem states
\begin{eqnarray}\label{simplifiedoptical}
  \ima \hat \Pi_{\mu\nu}
  =
  \frac{1}{2}
  \int \dd\Omega
  \;  \left(
  \frac{1-\cos p\theta q}{2}
  \right)
  A_{\mu\nu}.
\end{eqnarray}
We can immediately observe that, for space-time non commutativity,
unitarity is violated: we have $q \bullet q = \Theta_E
(p_0^2-p_1^2)+\Theta_B(p_2^2+p_3^2)<0$ for $q^\mu$ space-like, so
the left hand side of (\ref{simplifiedoptical}) is 
non-vanishing (because the argument of the Bessels becomes imaginary). At the
same time momentum conservation forces the right hand side to be
zero. This confirms the results in the literature.

Let us now focus on the case of purely space-like
non-commutativity, $\Theta_E=0$, which is more interesting. The
left hand side is easily evaluated from (\ref{pinchedPithetaanyd})
using
$$
\ima K_\nu(x)=(-1)^{\nu+1}\frac{\pi}{2} J_\nu(|x|).
$$
We find that for $q^2<0$ there are no imaginary parts, while for
$q^2>0$
\begin{eqnarray}
\label{imim}
\nonumber    \ima\widehat\Pi_{c}(q)
    &=&
    \frac{q^2}{2\pi}
    \left(
    \frac{11}{12}
    +
    \frac{
    2\sin z
    - 8 z^2\sin z
    - 2 z  \cos z
    }{ 8 z ^3}
    \right)
\\
    \ima\left\{\widehat\Pi_{c}(q)+\widehat\Pi_{\theta}(q)\right\}
    &=&
    \frac{q^2}{2\pi}
    \left(
    \frac{11}{12}
    +\frac{
    -4 \sin  z
    - 6 z^2  \sin z     + 4 z \cos z
    }{8 z^3}
    \right),
\end{eqnarray}
where $\displaystyle{ z=-\frac{|\tilde{q }| |q|}{2}}$. We consider
now the right hand side, that is more easily analyzed by employing
the transformations in the planes ($q^0,q^1$) and ($q^2,q^3$)
admitted by the residual Lorentz invariance: without loss of
generality we can take $q^\mu=(q^0,0,0,q^3)$ and we obtain
\[
  p\theta q
  =
  -\theta q^3 |\vec p| \cos\phi \sin \chi = -|\tilde{q }| |\vec p| \cos\phi \sin \chi
  ,
\]
where $(\chi,\phi)$ are polar and azimuthal angles defined with
respect to $x_3$. The phase space integrals evaluate to
\begin{eqnarray}
  &&\frac{1}{(2\pi)^6}
  \int\frac{\dd^3p\,\dd^3k}{4E_p E_k}
  \frac{(1-\cos p\theta q)}{2}
  (2\pi)^4\delta^{(4)}(p+q+k)
  A_{\mu\nu}
    =\nonumber\\
    &&=
  \frac{1}{8(2\pi)^2}
 \int_0^\pi \dd\theta
  \int_0^{2\pi} \dd\phi
  \,\sin\chi\,
  \left(\frac{1-\cos \left(-|\tilde{q }| |\vec p|  \cos\phi\sin\chi\right)}{2}\right)
  A_{\mu\nu}
  .
\end{eqnarray}
We can then  project  $A_{\mu\nu}$ on  the two independent
polarizations:
\begin{eqnarray}
  A^{\mu\nu}
   &\doteq&
  I_1 \left[g^{\mu\nu}-\frac{q^\mu q^\nu}{q^2}-\frac{\tilde{q }_\mu \tilde{q }_\nu}{|\tilde{q
  }|^2}\right]
  +I_2 \frac{\tilde{q }_\mu \tilde{q }_\nu}{|\tilde{q }|^2},
  \\
  I_1
  &=&
    \fract{7}{2}q^2
  - 2\frac{(p\theta q)^2}{|\tilde{q }|^2}=\fract{7}{2}q^2
  +2  \frac{|\tilde{q }|^2 |\vec p|^2 \cos\phi \sin \chi}{|\tilde{q
  }|^2},
  \\
  I_2
  &=&
   4q^2
  +4\frac{(p\theta q)^2}{|\tilde{q }|^2}=4q^2
  -4\frac{|\tilde{q }|^2 |\vec p|^2 \cos\phi \sin \chi}{|\tilde{q
  }|^2}
\end{eqnarray}
to obtain
\begin{eqnarray}
\nonumber   \int \dd\Omega
  \left(
  \frac{1-\cos p\theta q}{2}
  \right)
  I_1
  &=&
  \frac{q^2}{4\pi}
  \left(
  \frac{11}{12}
  +\frac{
  2 \sin z
  - 8 z^2 \sin z
  - 2z  \cos z
  }{8 z^3 }
  \right),
\\
  \int \dd\Omega
  \left(
  \frac{1-\cos p\theta q}{2}
  \right)
  I_2
  &=&
  \frac{q^2}{4\pi}
  \left(
  \frac{11}{12}
  +\frac{
  -4 \sin  z
  - 6  z^2  \sin z
  +z   \cos z
  }{ 8 z^3}
  \right).
\end{eqnarray}
This expression coincides  with one half of equation (\ref{imim}).
This completes our check of unitarity for the non-commutative
$U(1)$ gauge theory.

\subsection{Analyticity}

An analytic two-point function has the property that its real and
imaginary parts are related by a dispersion relation. Additional
care is needed when dealing with gauge field theories off-shell,
because of gauge artifacts. Unphysical gauge-dependent degrees of
freedom may introduce unphysical Landau poles that, in turn, can
spoil analyticity. The PT resummation prescription provides an
off-shell two-point function which takes only physical Landau
singularities into account. In the non-commutative setup we can
use this amplitude to investigate the deformation's effect on the
analytic structure of the two-point function. We start by checking
the dispersion relation involving the \textit{retarded
self-energy}'s real and imaginary parts
\begin{eqnarray}
\label{disp}
  \mathop{\rm Re} \Pi^{(\rm R)}(q)
  =
  \frac{1}{\pi}
  {\rm P}
  \int_{-\infty}^{+\infty}
  \dd \omega
  \frac{\mathop{\rm Im} \Pi^{(\rm R)}(\omega, \vec
  q)}{\omega-q^0}=\frac{2}{\pi}
  {\rm P}
  \int_{-\infty}^{+\infty}
  \dd \omega
  \omega \frac{\mathop{\rm Im} \Pi^{(\rm R)}(\omega, \vec
  q)}{\omega^2-(q^0)^2}.
\end{eqnarray}
As already pointed out in \cite{BDF2}, for positive energy, the
Feynman and retarded self-energies coincide, so an expression
analogous to (\ref{disp}) must hold for $\widehat \Pi_{\mu\nu}$.
Inserting equation (\ref{imim}) into (\ref{disp}) and evaluating
the integral we easily obtain the real part of the self-energy.
This, in turn, agrees with the value of the real part of the
self-energy extracted directly from (\ref{pinchedPithetaanyd}).
This shows that the two-point function's analyticity is preserved
in the non-commutative setup so the resummed amplitudes propagate
physically meaningful information.

The-non planar part of $\widehat \Pi^g_{\mu\nu}$
satisfies an \textit{unsubtracted dispersion relation} but  even
though the imaginary part is regular in the $\theta p \rightarrow
0$ limit it gives rise, once integrated, to an IR-divergent real
part.
 \FIGURE{
    \includegraphics[width=2.5in]{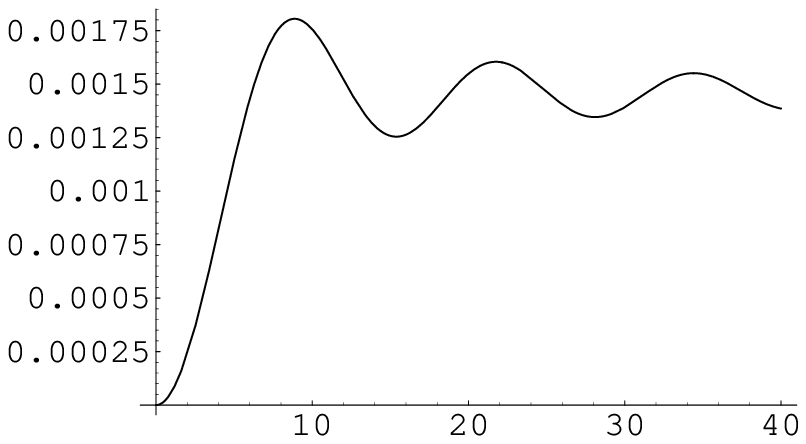}
    \includegraphics[width=2.5in]{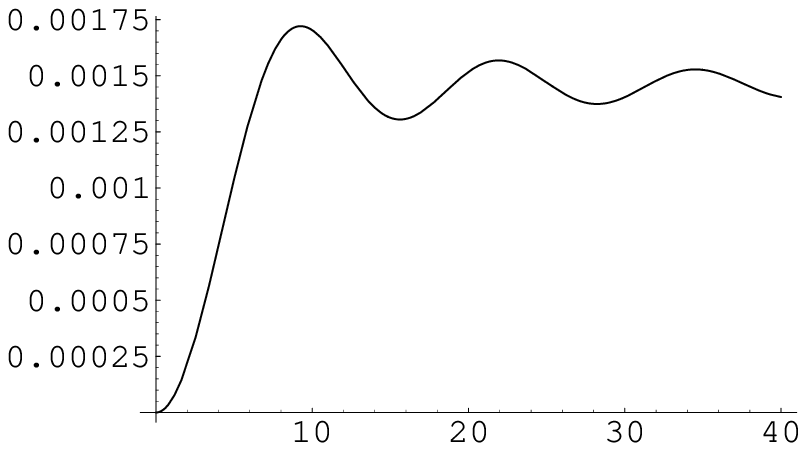}
  \caption{\label{imag1}
   The imaginary parts $\ima\widehat\Pi_{c}(q)$ and
  $\ima\left\{\widehat\Pi_c+\widehat\Pi_{\theta}\right\}$ as
  functions of $|\tilde{q}|$, for $g^2=10^{-2}$.}
  }
This divergence was observed in \cite{BDF2}, and it was
interpreted as an effect of the UV-IR mixing. However the
off-shell amplitudes considered there are gauge-dependent, so they
may contain unphysical degrees of freedom which in principle can
spoil the physical nature of this divergence. Working with the
pinched self-energy we proved that  the IR-divergence
comes out from the integration in the  high energy region and so it
is indeed a physical UV-IR effect.

\section{The two-dimensional theory}

In this section we will employ our gauge-invariant resummed
self-energy to analyze the two-dimensional theory. There are
several reasons to consider this apparently simple situation:
first of all, bidimensional non-commutative gauge theories present
a non-trivial behavior at the perturbative level, exhibiting
unexpected effects when Wilson loops are evaluated. It was found
in \cite{bacc1} that, in the large-$\theta$ limit, non-planar
contributions are not suppressed, leading to an anomalous finite
result. Later on, it was shown \cite{mak} that area-preserving
diffeomorphism invariance is violated in perturbation theory, a
surprising feature further confirmed by recent computations
\cite{bacc2} (see also \cite{rick} for a nonperturbative approach
to this problem).

Since all these phenomena appear at the perturbative level in
gauge-invariant observables, it is quite tempting to explore the
gauge-invariant  propagator itself. Actually all of
these results were obtained by using axial gauges, that explicitly
trivialize the self-interactions of the gauge fields in
two-dimensions. This procedure leads to infrared-finite results,
without resorting to the choice of any explicit cut-off once a
suitable prescription for the gauge propagator is adopted. In our
case, instead, we have used a covariant gauge-fixing in deriving
the self-energy in $D$-dimensions: by taking the limit
$D\rightarrow 2$ in our general expression, we will obtain
automatically a dimensionally regularized result.

In both the $D\to 2$ and $\theta\to 0$ limits one would naively
expect a vanishing gauge-invariant self-energy when matter is
absent: two-dimensional gauge theories have no physical local
degrees of freedom and, since we are considering the $U(1)$-case,
a free theory is recovered for $\theta=0$ at the classical level.
This last feature should be true even in presence of matter. We
anticipate that quantum effects will produce a quite different
behavior  instead, as we will see in the following.

\subsection{The gluon contribution}

We start by recalling the expression for $\widehat\Pi^g_{\mu\nu}$
in equation (\ref{integratedpinchedPianyd}):
\begin{eqnarray}
  \widehat\Pi^g_{\mu\nu}(q)
  &=&
  \frac{-g^2}{(4\pi)^{D/2}} \int_0^1\dd x \Bigg\{
  \frac{8q^2+(4-2D)(q^2x^2-M_g^2)}{(M_g^2)^{2-D/2}}\left(g_{\mu\nu}-\frac{ q_\mu q_\nu}{q^2}\right)
  \times\nonumber\\&
  \times&\left[\Gamma(2-D/2)-2 \left(\frac{|\tilde q| M_g}{2}\right)^{2-D/2} K_{2-D/2}(M_g |\tilde{q}|)\right]
  +\nonumber\\&+&
  (4-2D)\Bigg[2 \left(\frac{1}{2M_g}\right)^{1-D/2} \left(|\tilde q|M_g\right)^{2-D/2}K_{-D/2}(M_g |\tilde{q}|)\Bigg]
  \frac{\tilde q_\mu \tilde q_\nu}{\tilde q^2} \Bigg\}.
  \label{2d}
\end{eqnarray}
We notice that for $2<D<4$ the limit $\theta\to 0$ can be taken
safely in the first term, obtaining the expected decoupling in the
usual transverse part, while the new transverse structure produces
the well known $1/\theta^{(D-2)}$ divergence, generated by the
IR/UV mixing. On the other hand, as $\theta\to\infty$ we recover
the pure planar theory.

Things change when one tries to go down to $D=2$. A first
observation concerns the infrared and ultraviolet behavior. In
order to understand the potential divergences and the peculiar
role played by non-commutativity, it is useful to take a step back
and look directly at the Feynman integral
\begin{equation}
  \int d^D k\Biggl[\frac{ 8q^2 g_{\mu \nu}}{k^2 (q+k)^2}+(4-2D)\frac{(k^2 g_{\mu \nu}
  -2 k_\mu k_\nu) }{ k^2 (q+k)^2} \Biggr]\sin^2\left(\frac{q \theta  k}{2}\right).
  \label{2d11}
\end{equation}
 One immediately sees how non-commutativity plays a
crucial role in approaching two dimensions: the potential infrared
divergence in the first term is smoothed out by the
$\sin^2(\frac{q \theta  k}{2})$ term. Here non-commutativity acts
as an infrared regulator when one computes the momentum integral.
The second contribution is instead a classical example of an
\textit{evanescent term}, being multiplied by $(4-2D)$, that in
presence of ultraviolet and infrared divergences could generate a
finite result in the limit $D\rightarrow 2$.

A phenomenon of this type was noticed in \cite{bacci}, where
Wilson loops were computed in non-abelian $D=2$ gauge theories by
using dimensional regularization. In our case non-commutativity
provides a natural cut-off and the evanescent term  gives no
contribution in the two-dimensional limit. These features are
displayed by explicitly performing the limit $D\rightarrow 2$ in
equation (\ref{2d}): we simply obtain
\begin{equation}
  \widehat\Pi^g_{\mu\nu}(q)= -\frac{ g^2}{(4\pi)}\Bigg[\int_0^1 \dd
  x \frac{\left( 8q^2\right)}{M^2}
  \Bigg(1-  \left(|\tilde q| M_g\right) K_{1}(M_g |\tilde{q}|) \Bigg)
  \Bigg](g_{\mu\nu}-\frac{ q_\mu q_\nu}{ q^2}).
  \label{2d1}
\end{equation}
Here one can see the role played by non-commutativity: we get an
infrared-regulated contribution from the first term of (\ref{2d}),
which owes its finiteness to a delicate cancellation between the
planar and the non-planar sectors, piloted by the non-commutative
phase. We
remark that the term proportional to
$\tilde q_\mu \tilde q_\nu$  in equation (\ref{2d}), which displayed in $D=4$ the 
known  IR/UV effects,  is the product of a finite term times $D-2$ and so it vanishes.

Now we can explore a couple of different limits. Using the
expansion of the Bessel function we take the limit $s=(|\tilde
q||q|/2) \rightarrow 0$, observing that the full self-energy
vanishes. We recover the free theory result, as could be well
expected because no ultraviolet divergence was regularized by
non-commutativity in the final expression. We will see that when
matter is present the situation changes drastically. The opposite,
$s\rightarrow \infty$ limit is more interesting: this computation
should reproduce the planar part of a commutative non-abelian
theory. In order to perform the calculation, we write down the
relevant part of the self-energy as follows:
\begin{eqnarray}
  \int_0^1 \dd x &&\!\!\!\!\!\!\!\frac{ q^2}{M_g^2}
  \left[1-  \left(|\tilde q| M_g\right) K_{1}(M_g|\tilde{q}|) \right]
  =2\int_0^1 \dd x \frac{1}{\sqrt{1-x}}\left[ \frac{1}{x}- (|\tilde
  q||q|/2)\frac{1}{\sqrt{x}} K_1(|\tilde q||q|\sqrt{x}/2 )
  \right]\!\!
  =\nonumber\\ &=&
  2\int_0^1 \dd x
  \left[\frac{1}{\sqrt{1-x}}\frac{1}{x}-\frac{K_1( \sqrt{x}
  )}{\sqrt{x(1-x/s^2)}} +\frac{K_1( \sqrt{x} )}{\sqrt{x(1-x/s^2)}} -
  \frac{2 s K_1(s \sqrt{x} )}{\sqrt{x(1-x)}} \ \right].
\end{eqnarray}
The above expression can be easily evaluated in the limit
$s\rightarrow \infty$:
\begin{eqnarray}
  2\left[\int_0^1 \dd x \left(
  \frac{1}{\sqrt{1-x}}\frac{1}{x}-
  \frac{K_1( \sqrt{x} ) }{\sqrt{x}}\right) +\ln(s^2)+\int_1^\infty
  \dd x \frac{K_1( \sqrt{x} ) }{\sqrt{x}}
   \right]
   \rightarrow 2\ln(s^2),
\end{eqnarray}
implying the following behavior of the vacuum polarization
\begin{equation}
  \lim_{s\to\infty}\widehat\Pi^g_{\mu\nu}=
      -g^2\frac{8}{\pi
      }\ln(s)\left(g_{\mu\nu}-\frac{ q_\mu q_\nu}{q^2}\right).
\end{equation}
We observe therefore a curious ``twisted" incarnation of the
familiar IR/UV mixing: the original infrared divergence is cured
by the non-commutativity, and it reappears as an ultraviolet
effect as $s\to \infty$. In this limit the effective infrared
cut-off is removed, since the non-planar contribution is
suppressed completely. We remark that our result is a simple
example of how, in two dimensions, the limit of large-$\theta$ can
produce non-trivial effects in perturbation theory, as shown in
\cite{bacc1,mak} by computing Wilson loops.

\subsection{ Fermionic and scalar contributions}

It is a simple exercise to compute the contribution to self-energy
of $n_f$ fermions and $n_s$ scalars, taking the limit $D=2$ in the
general expressions (\ref{susypi}) and  (\ref{susypi1}):
\begin{eqnarray}
  \widehat\Pi^f_{\mu\nu}    &=&
     \sum_{n_f} \frac{g^2}{\pi}
     \int_0^1 \dd x
       \Bigg\{
      \frac{q^2 x(1-x)}{M_f^2}
      \left[1-  \left(|\tilde q| M_f\right) K_{1} \right]
        +    \left(M_f |\tilde q|\right) K_{1}
                     \Bigg\}(g_{\mu\nu}-\frac{ q_\mu
                     q_\nu}{q^2}),
\\
  \widehat\Pi^s_{\mu\nu}
  &=&
  \sum_{n_s}   \frac{g^2}{\pi}
     \int_0^1 \dd x
    \Bigg\{
       \frac{q^2 (x^2-1/4)}{M_s^2}
      \left[1-  \left(|\tilde q| M_s\right) K_{1} \right]
  - \left( M_s|\tilde q| \right) K_{1 }    \Bigg\}  (g_{\mu\nu}-\frac{
  q_\mu q_\nu}{q^2}).
\end{eqnarray}
The above expressions are finite even in the massless case: in
particular the fermionic contribution is regular independently of
the cancellation between planar and non-planar sectors, and the
infrared safeness of the massless scalar integral is ensured by
the same mechanism as in the gauged case.

We observe, at variance with the pure gauge case, that the
structure ${\tilde q_\mu \tilde q_\nu}/{\tilde q^2}$ leads to a
finite contribution. Since in two dimensions
$\theta_{\mu\nu}=\theta \epsilon_{\mu\nu}$, it happens that
${\tilde q_\mu \tilde q_\nu}/{\tilde q^2}=g_{\mu \nu}-{ q_\mu
q_\nu}/{ q^2}$, and no new structure appears. Taking the limit
$s\rightarrow 0$ we have the following surprising result:
\begin{eqnarray}
  \widehat\Pi^f_{\mu\nu}  \to
  \sum_{n_f} \frac{g^2}{(\pi)} \int_0^1 \dd x \left(M_f |\tilde
  q|\right) K_{1}(M_f |\tilde{q}|) (g_{\mu\nu}-\frac{ q_\mu q_\nu}{q^2})\to n_f
  \frac{g^2}{\pi}(g_{\mu\nu}-\frac{ q_\mu q_\nu}{q^2}),
\\
  \widehat\Pi^s_{\mu\nu}\to
  -\sum_{n_s}   \frac{g^2}{\pi}\int_0^1 \dd x    \left( M_f|\tilde q|
  \right) K_{1 }(M_f |\tilde{q}|) (g_{\mu\nu}-\frac{ q_\mu q_\nu}{q^2})\to -n_s
  \frac{g^2}{\pi}(g_{\mu\nu}-\frac{ q_\mu q_\nu}{q^2}).
\end{eqnarray}
We are left with a constant term which wastes the decoupling: this
is a ``canonical" IR/UV effect, as one can easily realize, because
it originates from the ${\tilde q_\mu \tilde q_\nu}/{\tilde q^2}$
term. For $D>2$ it produces the well known $1/ \theta^{(D-2)}$
divergence, while in two dimensions it provides a finite,
$\theta$-independent result as the non-commutativity is sent to
zero.

This effect of producing a Schwinger mass at one-loop for the
gauge field is the exact analogue of the induction of a
Chern-Simons term in three dimension when Majorana fermions are
coupled to a non-commutative $U(1)$ theory \cite{chu}. In that
case too a non-vanishing Chern-Simons term, generated by the
non-commutative interaction, survives as $\theta\to 0$, leading to
a one-loop mass for the gauge field.

The opposite situation, the limit $s \to \infty$ is
straightforward to compute in the massive case, the Bessel
function being exponentially suppressed, and we are left with the
regular planar contributions. As we anticipated before, the
massless scalar case exhibits the same anomalous behavior as the
pure gauge case at large $s$: exploiting the same technique, we
easily derive
\begin{equation}
  \lim_{s\to \infty}\widehat\Pi^s_{\mu\nu}=
      \frac{n_s}{3}\frac{g^2}{\pi
      }\ln(s)(g_{\mu\nu}-\frac{ q_\mu q_\nu}{q^2}).
\end{equation}
When $n_s=24$ this cancels exactly the anomalous divergence coming
from the gauge sector. We do not have an  an explanation for this
curious fine-tuning.

\section{Conclusions and outlook}

An interesting issue in non-commutative gauge theories is whether
unitarity and analyticity of the Green functions are spoiled by
non-local effects, both on-shell and off-shell. A consistent,
gauge-invariant resummation formalism is required to investigate
these properties, and a leading candidate is provided by the pinch
techniques framework.

In this paper we have worked out a gauge-invariant resummation
prescription, extending the pinch techniques to non-commutative
gauge theories. We have shown how resummed off-shell Green's
functions satisfy the requirements that are usually imposed in the
commutative setup to implement a consistent gauge-invariant
reorganization of the perturbative expansion. In particular, an
important check of the validity of the pinch techniques requires
that the resummed self-energy reduce on-shell to the unpinched
one. We have verified that this is the case: in particular, the
pinched gluon self-energy displays, in four dimensions, the well
know tachyonic divergence that leads to vacuum destabilization.

With this gauge-invariant  resummation formalism at our disposal
we proceeded to carry out an analysis of the optical theorem,
verifying that the pinch techniques provide a powerful tool for
investigating the issue of unitarity of non-commutative gauge
theories. Previous analyses in this field employed techniques
which were sensitive to unphysical gauge effects; our main result
is a test of the optical theorem in the $s$-channel, employing
off-shell resummed functions, for purely spacelike
non-commutativity. For timelike non-commutativity we found
evidence for unitarity violation both on-shell and off-shell,
consistently with previously known results. 

Finally we came to the analysis of the $D \rightarrow 2$ limit.
The two-dimensional theory is expected to be trivial due to the
absence of propagating degrees of freedom for the pure gauge sector.
We found instead  a non-trivial correction to the dispersion
relation even in the absence of matter. Moreover, when matter is
included we found an anomalous behavior in the $\theta
\rightarrow 0 $ limit. A finite term survives the commutative
limit, violating the expected decoupling and inducing a mass term
for the photon. In the $\theta \rightarrow \infty$ limit instead,
a twisted version of the UV-IR mixing comes out: the original
infrared divergences regulated by the non-commutativity reappear
in the ultraviolet domain.

In the longer run, the most interesting issues are related with
the possibility of writing down a consistent Schwinger--Dyson
equation to investigate vacuum destabilization from a
non-perturbative point of view. There are at least two motivations
for doing so. A first reason is related to the possibility that
vacuum destabilization might simply be an artifact of perturbation
theory. A more speculative motivation is related to the
possibility for the existence of striped phases like the ones
observed in non-commutative scalar theories \cite{gubso} in the
gauged case. It should be remarked that a transition to a striped
phase here would be particularly puzzling, since translations form
a subset of the full $U(1)_\star$ gauge group. Quite surprisingly,
however, we
found hints that such an exotic phase might be realized in the
three-dimensional topologically massive case
\cite{Caporaso:2004hq,TMYMCS}.

For this purpose, the next step would be to write down a
gauge-invariant gap equation for the pinch-technique propagator
$\widehat \Delta$. Truncating the Schwinger--Dyson equations to
one-loop order one has
\begin{eqnarray}\label{gap_equation}
  \hat\Delta^{-1}(q)
  &=&
  \hat\Delta_0^{-1}(q)
  +\widehat V^{(3)}(q,p,k) \widehat \Delta(p) \widehat \Delta(k) \widehat V^{(3)}(-k,-p,-q)
  + \nonumber \\&&
  +\widehat V^{(4)}(q,p,-q,-p) \widehat \Delta(p)
  +\mbox{pinch terms}
  ,
\end{eqnarray}
where $\widehat V$ are the full vertices. The pinch terms are the
usual ones, but they must be computed using the exact
pinch-technique propagators $\widehat\Delta(q)$, for which one can
take an Ansatz of the form
\begin{equation}\label{gap_propagator}
  \hat\Delta(q)
  =
  T(q)\left(-g_{\mu\nu}+\frac{q_\mu q_\nu}{q^2}\right)
  +\Theta(q)\left(\frac{\tilde{ q}_\mu \tilde{ q}_\nu}{q^2}\right)
  +(1-\xi)\left(\frac{q_\mu q_\nu}{q^4}\right) .
\end{equation}
The propagator's trivial dependence on the gauge fixing is
retained in the framework of the pinch techniques, as explained
above, and this should be used to test that the truncation is
indeed self-consistent and gauge-independent.

A first trivial attempt consists, for example, in seeing what
happens when one computes the gap equation \eqref{gap_equation}
for non-commutative QCD, with all pinch terms included, using the
tree-level form of the vertices. So doing one encounters
encouraging cancellations among the $\xi$-dependent terms, but
there are some gauge-fixing-dependent terms (\textit{e.g.} those
proportional to $\Theta(q)$) that don't cancel. This is not at all
a surprise, since one should in principle determine the $\widehat
V^{(3)}$ vertex through its own Scwhinger--Dyson equation. In the
commutative case one can use the pinch-technique Ward identities
\begin{equation}
  q_\alpha
  (\widehat V-\widehat V_0)_{\alpha\mu\nu}(q,p,-q,-p)
  =
  \widehat \Pi_{\mu\nu}(q+p)
  -\widehat \Pi_{\mu\nu}(p)
\end{equation}
to find the form of $\widehat V^{(3)}$. Hopefully an analogous
approach can lead, like in the commutative case, to a meaningful,
gauge-invariant gap equation.

\bigskip
\noindent
\textbf{Acknowledgements}
We would like to thank L. Griguolo and D. Seminara 
for useful discussions
during the making of this paper, as well as for carefully reading
this manuscript.

\newpage
\appendix
{\Large \bf Appendices}\\
\addcontentsline{toc}{section}{Appendices}
\section{Feynman Rules}
In this appendix we shall summarize the Euclidean Feynman rules
adopted in this paper

{\it \textsc{Gluon sector:}} \vspace{-2cm}
\begin{center}
{
\begin{picture}(400,100)
\Gluon(10,20)(100,20){4}{10} \ArrowLine(40,10)(60,10)
\Text(18,32)[]{$\mu$}\Text(92,32)[]{$\nu$}\Text(55,32)[]{$p$}
\Text(220,20)[]{$\displaystyle{\frac{1}{p^2}\left(g_{\mu\nu}-(1-\xi)p_\mu
p_\nu\right)}$}
\end{picture}}
\end{center}
\begin{center}
{
\begin{picture}(400,100)
\Gluon(50,100)(50,50){4}{6} \ArrowLine(58,85)(58,65)
\ArrowLine(86,24)(70,40) \ArrowLine(15,24)(31,40)
\Gluon(50,50)(10,10){4}{6}\Gluon(50,50)(90,10){4}{6}
\Text(5,20)[]{$\mu$}\Text(28,12)[]{$q$}
\Text(95,20)[]{$\nu$}\Text(75,12)[]{$r$}
\Text(60,90)[]{$\lambda$}\Text(40,90)[]{$p$}
\Text(240,60)[]{$-(2\pi)^d\delta(p+q+r)2i
g(g_{\lambda\mu}(p-q)_\nu+g_{\mu\nu}(q-r)_\lambda+$}
\Text(280,40)[]{$+g_{\nu\lambda}(r-p)_\mu)\sin\left(\frac{p\theta
q}{2}\right)$}
\end{picture}}
\end{center}
\begin{center}
{
\begin{picture}(400,100)
\Gluon(50,50)(90,90){4}{6}
\Gluon(50,50)(90,90){4}{6}\Gluon(50,50)(10,90){4}{6}
 \ArrowLine(19,70)(35,54)
\ArrowLine(86,24)(70,40)
\ArrowLine(15,24)(31,40)\ArrowLine(80,70)(64,54)
\Gluon(50,50)(10,10){4}{6}\Gluon(50,50)(90,10){4}{6}
\Text(5,20)[]{$\nu$}\Text(28,12)[]{$q$}
\Text(95,20)[]{$\alpha$}\Text(75,12)[]{$r$}
\Text(23,90)[]{$\mu$}\Text(5,83)[]{$p$}\Text(78,90)[]{$s$}\Text(94,83)[]{$\beta$}
\Text(230,50)[]{$(2\pi)^d\delta(p+q+r+s) (-4
g^2)\biggl(\sin\frac{p\theta q}{2}\sin\frac{r\theta
s}{2}(g_{\mu\alpha}g_{\nu\beta}-g_{\mu\beta}g_{\nu\alpha})+$}
\Text(260,30)[]{$+ \sin\frac{p\theta r}{2}\sin\frac{q\theta
s}{2}(g_{\mu\nu}g_{\alpha\beta}-g_{\mu\beta}g_{\nu\alpha})+$}
\Text(260,10)[]{$+ \sin\frac{q\theta r}{2}\sin\frac{p\theta
s}{2}(g_{\mu \nu}g_{\alpha\beta}-g_{\nu\beta}g_{\mu\alpha})
\biggr)$}
\end{picture}}
\end{center}
{\it \textsc{Ghost sector:}}
\begin{center}
{\vspace{-2.5cm}
\begin{picture}(400,100)
\SetWidth{1}
\Line(20,20)(120,20)\SetWidth{.5}\ArrowLine(50,15)(90,15)\Text(250,20)[]{$\displaystyle{\frac{1}{p^2}
}$} \Text(70,8)[]{$p$}
\end{picture}}
\end{center}
\begin{center}
{
\begin{picture}(400,100)
\SetWidth{1}\ArrowLine(70,10)(100,40)\ArrowLine(100,40)(130,10)
\SetWidth{.5}\ArrowLine(130,20)(110,40)\ArrowLine(70,20)(90,40)
\Text(250,40)[]{$\displaystyle{(2\pi)^d\delta(p+q+r) 2 i g p^\mu
\sin\frac{p\theta q}{2}}$}\SetWidth{.5}\Text(140,20)[]{$p$}
\Gluon(100,40)(100,85){4}{5}\Text(110,80)[]{$\mu$}\ArrowLine(110,70)(110,55)\Text(90,62)[]{$q$}
\end{picture}}
\end{center}
{\it \textsc{Dirac Fermions:}}
\begin{center}
{\vspace{-2.5cm}
\begin{picture}(400,100)
\SetWidth{1}
\Line(20,20)(120,20)\SetWidth{.5}\ArrowLine(50,15)(90,15)\Text(250,20)[]{$\displaystyle{\frac{1}{\ssh
p+m} }$} \Text(70,8)[]{$p$}
\end{picture}}
\begin{center}
{
\begin{picture}(400,100)
\SetWidth{1}\ArrowLine(70,10)(100,40)\ArrowLine(100,40)(130,10)
\SetWidth{.5}\ArrowLine(130,20)(110,40)\ArrowLine(70,20)(90,40)
\Text(250,40)[]{$\displaystyle{(2\pi)^d\delta(p+q+r)  g e^{\frac{i
r\theta p}{2}}}$}\SetWidth{.5}\Text(140,20)[]{$p$}
\Text(60,20)[]{$r$}
\Gluon(100,40)(100,85){4}{5}\Text(110,80)[]{$\mu$}\ArrowLine(110,70)(110,55)
\Text(90,62)[]{$q$}
\end{picture}}
\end{center}
\end{center}
{\it \textsc{Maiorana Fermions:}}
\begin{center}
{\vspace{-2.5cm}
\begin{picture}(400,100)
\SetWidth{1}
\Line(20,20)(120,20)\SetWidth{.5}\ArrowLine(50,15)(90,15)\Text(250,20)[]{$\displaystyle{\frac{1}{\ssh
p+m} }$} \Text(70,8)[]{$p$}
\end{picture}}
\begin{center}
{
\begin{picture}(400,100)
\SetWidth{1}\ArrowLine(70,10)(100,40)\ArrowLine(100,40)(130,10)
\SetWidth{.5}\ArrowLine(130,20)(110,40)\ArrowLine(70,20)(90,40)
\Text(250,40)[]{$\displaystyle{(2\pi)^d\delta(p+q+r)  g
\sin\left({\frac{r\theta
p}{2}}\right)}$}\SetWidth{.5}\Text(140,20)[]{$p$}
\Text(60,20)[]{$r$}
\Gluon(100,40)(100,85){4}{5}\Text(110,80)[]{$\mu$}\ArrowLine(110,70)(110,55)
\Text(90,62)[]{$q$}
\end{picture}}
\end{center}
\end{center}
Where $p \theta q=p_\mu \theta^{\mu\nu}q_\nu$.
\section{Relevant scalar and  euclidean tensorial integrals}
This appendix is devoted to the evaluation of the relevant scalar
and tensorial integrals. To begin with, we shall consider the
tadpole-like integral, which is taken to be  identically zero in
the commutative case, \beq \label{Int1} T= \int \frac{d^d
k}{(2\pi)^d} \sin^2 \left(\frac{k\theta q}{2}\right)
\frac{1}{k^2}. \eeq This integral for $d\ge 2$ is ultraviolet
divergent, but infrared finite. In fact the presence of the
trigonometric function  smooths the behavior for small momenta.
This should be contrasted with the commutative counterpart where
dangerous infrared divergences appear when $d$ approaches two.

A rigorous dimensional regularization  of  the integral
(\ref{Int1})  requires its evaluation for $d<2$ and then to define
its values in $d\ge 2$ by analytic continuation. We have \bea
T\!\!&=&\!\!\frac{1}{2}\lim_{M\to 0} \int_0^\infty \!\!\!\! dt\!\!
\int \!\!\frac{d^d k}{(2\pi)^d} \left(\!1-e^{- i{k\theta
q}}\!\right)\! e^{-t (k^2+M^2)}\!\!=
\frac{\pi^{d/2}}{2(2\pi)^d}\lim_{M\to 0} \int_0^\infty \!\!\!\!
\frac{dt}{t^{d/2}}\!\!  \left(1-e^{-\frac{|\tilde{q}|^2 }{4 t}}\right) e^{-t M^2}\!\!\!\!\!=\nonumber\\
\!\!&=&\!\! \frac{1}{2(4\pi)^{d/2}}\lim_{M\to 0}\left(
\Gamma\left(\frac{d}{2}-1\right)
(M^2)^{d/2-1}-2^{\frac{d}{2}}{\left( M^2 \right) }^
   {\frac{d}{2}-1}
  {\left( M^2|\tilde{q}|^2 \right) }^
   {\frac{1}{2} - \frac{d}{4}}
  \Mfunction{K}_{d/2 -1}(M |\tilde{q}| )\right)\!=\nonumber\\
   &=&\frac{1}{2(4\pi)^{d/2}}\lim_{M\to 0}\left[
   \left(
    2^{ d-4}|\tilde{q}|^{4 - d}\Mfunction{\Gamma}
    \left(\frac{d-4 }{2}\right){\Mfunction{M}}^2
   - 2^{d-2}|\tilde{q}|^{2 - d}\Mfunction{\Gamma}\left( \frac{d}{2}-1 \right)   +
   {\Mfunction{O}(M^4)}
    \right)  +\right.\nonumber\\
   \!\!&+&\!\!\left. M^d\left( \frac{|\tilde{q}|^2}{4}\Mfunction{\Gamma}\left(-\displaystyle{\frac{d}{2}}
   \right) -
     \frac{|\tilde{q}|^4}{32}\Mfunction{\Gamma}\left(
     -1 - \displaystyle{\frac{d}{2}}\right){\Mfunction{M}}^2 + {\Mfunction{O}(M^3)}
      \right)
    \right]\!\!\!=-\frac{2^{d-3}|\tilde{q}|^{2 - d}}{(4\pi)^{d/2}}
    \Mfunction{\Gamma}\left(\frac{d}{2}-1\right)\!\!.
\eea Around $d=2$, we have the following expansion \beq T=
-\frac{1}{4\,\pi \,\left( d -2\right) } +
   \frac{\Mvariable{\gamma}}{8\,\pi } + \frac{\log (\pi\mu^2|\tilde{q}|^2
     )}{8\,\pi }  +
  \Mfunction{O}(d-2).
\eeq The second scalar integral we need is given by \bea S&=&\int
\frac{d^d k}{(2\pi)^d}\frac{1}{k^2(k+q)^2} \sin^2
\left(\frac{q \theta k}{2}\right)=\nonumber\\
&=&\frac{1}{2}\Biggl[ \int \frac{d^d
k}{(2\pi)^d}\underbrace{\frac{1}{k^2(k+q)^2}}_{(a)}- \int
\frac{d^d k}{(2\pi)^d}\underbrace{\frac{1}{k^2(k+q)^2}e^{i k
\theta q}}_{(b)} \ \Biggr]. \eea We have \beq
S_{(a)}=\frac{1}{2}\int_0^1 dx \int\frac{d^d
k}{(2\pi)^d}\frac{1}{(k^2+x(1-x)q^2)^2}= \frac{\Gamma(2-d/2)}{2
(4\pi)^{d/2}}\int_0^1 dx (x(1-x)q^2)^{d/2-2}. \eeq and \bea
S_{(b)}&=&\frac{1}{2}\int_0^1 dx \int\frac{d^d
k}{(2\pi)^d}\frac{e^{i k\theta q}}{(k^2+x(1-x)q^2)^2}
=\frac{1}{2}\int_0^1\!\!\!\! dx\!\!\int_0^\infty\!\!\!\! \!dt\,
t\!\! \int\!\!\! \frac{d^d k}{(2\pi)^d}
e^{i k\theta q-t (k^2+x(1-x) q^2) }=\nonumber\\
&=&
\frac{1}{2(4\pi)^{d/2}}\int_0^1\!\!\!\!
dx\!\!\int_0^\infty\!\!\!\! dt \ t^{1-d/2}
e^{-\frac{|\tilde{q}|^2}{4t}-t x(1-x) q^2 }=\frac{1}{(4\pi)^{d/2}}
\,{\left( \frac{|\tilde{q}| }{2 M} \right)  }^{2-\frac{d}{2}}\,
 \!\!\!\!\!\!\! {K}_{\frac{d}{2}-2}(|\tilde{q}|  M),
\eea with $ M={\sqrt{   q^2 x\left( 1 - x \right) } }$. Next we
consider the tensorial integral \beq I_{\mu\nu}=\int \frac{d^d
k}{(2\pi)^d}\frac{(2 k_\mu +q_\mu)(2 k_\nu + q_\nu)-2 k^2
g_{\mu\nu}}{k^2(k+q)^2} \sin^2 \left(\frac{q\theta k}{2}\right).
\eeq This is manifestly transverse: in fact
 \bea
   q^\mu
   I_{\mu\nu}&=&\int \frac{d^d k}{(2\pi)^d}\frac{(2 (k\cdot q)
   +q^2)(2 k_\nu + q_\nu)-2 k^2 q_\nu }{k^2(k+q)^2}
   \sin^2 \left(\frac{q\theta k}{2}\right)=\nonumber\\
   &=&\int \frac{d^d k}{(2\pi)^d}\frac{((q+k)^2-k^2)(2 k_\nu +
   q_\nu)-2 k^2 q_\nu }{k^2(k+q)^2}
   \sin^2 \left(\frac{q\theta k}{2}\right)=\nonumber\\
   &=&\int \frac{d^d k}{(2\pi)^d}\left(\frac{(2 k_\nu +
   q_\nu)}{k^2}-\frac{(2 k_\nu + q_\nu)}{(k+q)^2}- 2 \frac{ q_\nu
   }{(k+q)^2}\right)
   \sin^2 \left(\frac{q\theta k}{2}\right)=\nonumber\\
   &=&\int \frac{d^d k}{(2\pi)^d}\left(\frac{ q_\nu}{k^2}+\frac{
   q_\nu}{k^2}- 2 \frac{ q_\nu }{k^2}\right) \sin^2
   \left(\frac{q\theta k}{2}\right)=0.
 \eea
To compute this integral, first we decompose it in its planar and
non planar parts
 \bea
   I^{\mathrm{plan.}}_{\mu\nu}&=&\frac{1}{2}\int \frac{d^d k}{(2\pi)^d}\frac{(2 k_\mu +q_\mu)(2 k_\nu + q_\nu)-2 k^2
   g_{\mu\nu}}{k^2(k+q)^2}
   \nonumber\\
   I^{\mathrm{non-plan.}}_{\mu\nu}&=&\frac{1}{2}\int \frac{d^d k}{(2\pi)^d}\frac{(2 k_\mu +q_\mu)(2 k_\nu + q_\nu)-2 k^2
   g_{\mu\nu}}{k^2(k+q)^2}\ e^{i k\theta q}
 \eea
and then we compute them separately. To avoid the question about
how to extend the different tensor structures in non integer
dimensions, we shall compute the integrals following the most
straightforward path, which always begins by introducing the
Feynman parameters \bea \label{pi9}
I^{\mathrm{plan.}}_{\mu\nu}&=&\frac{1}{2}\int_0^1 dx\int \frac{d^d
k}{(2\pi)^d}\frac{(2 k_\mu +(1-2 x)q_\mu)(2 k_\nu +(1-2 x)
q_\nu)-2 (k-x q)^2 g_{\mu\nu}}
{(k^2+x(1-x) q^2)^2}=\nonumber\\
&=&\frac{1}{2}\int_0^1 dx\int \frac{d^d k}{(2\pi)^d} \frac{2((2-d)
k^2/d -x^2 q^2) g_{\mu\nu}+(1-2 x)^2 q_\mu q_\nu}{(k^2+x(1-x)
q^2)^2}. \eea By employing the following basic result of
dimensional regularization, \beq \int \frac{d^d k}{(2\pi)^d}
\frac{(k^2)^\nu}{(k^2+M^2)^\mu}=\frac{(M^2)^{\nu-\mu+d/2}}{(4\pi)^{d/2}}\frac{\Gamma(\nu+d/2)
\Gamma(\mu-\nu-d/2)}{\Gamma(d/2)\Gamma(\mu)}. \eeq we can evaluate
all the integral over momenta in eq. (\ref{pi9}) \bea \label{pi10}
I^{\mathrm{plan.}}_{\mu\nu}&=&\frac{1}{2(4\pi)^{d/2}}\int_0^1
dx\left[\frac{2(2-d)}{d}g_{\mu\nu} (x(1-x)q^2)^{d/2-1}
\frac{\Gamma(1+d/2)
\Gamma(1-d/2)}{\Gamma(d/2)\Gamma(2)}+\right.\nonumber\\
&&\left.+((1-2x)^2 q_\mu q_\nu-2 x^2
q^2g_{\mu\nu}){(x(1-x)q^2)^{d/2-2}}\frac{
\Gamma(2-d/2)}{\Gamma(2)}\right]=\nonumber\\
&=&\frac{1 }{2(4\pi)^{d/2}} \left(q_\mu q_\nu-q^2 g_{\mu\nu}
\right)\int_0^1 dx\frac{\Gamma(2-d/2)}{(M^2)^{2-d/2}} (1-2x)^2.
      \eea
The term linear in $1-2 x$ vanishes because it is a total
derivative.

 The non-planar contribution can be computed by means of
the same techniques. The computation is however more tedious
because of the presence of the new vector $\tilde{q}^\mu$. \bea
\label{pi8}
I^{\mathrm{n-pl.}}_{\mu\nu}&=&\frac{1}{2}\int_0^1 dx\int \frac{d^d k}{(2\pi)^d}\frac{4 k_\mu
k_\nu +(1-2 x)^2 q_\mu q_\nu-2 (k^2+x^2 q^2) g_{\mu\nu}}
{(k^2+x(1-x) q^2)^2}e^{i k\theta q}
=\nonumber\\
&=&\!\!\frac{1}{2}\int_0^1\!\!\!\! dx\!\!\int_0^\infty\!\!\!\!
\!dt\, t\!\! \int\!\!\! \frac{d^d k}{(2\pi)^d}\left[{4 k_\mu
k_\nu\!\! +\!\!(1\!-\!2 x)^2 q_\mu q_\nu\!-\!2 (k^2+x^2 q^2)
g_{\mu\nu}}\right]
e^{i k\theta q-t (k^2+x(1-x) q^2) }=\nonumber\\
&=&\frac{1}{2}\frac{1}{(4\pi)^{d/2}}\int_0^1 dx\int_0^\infty dt\,
t^{1-d/2} \left[4 \left(\frac{g_{\mu\nu}}{2 t}-\frac{\tilde{q}_\mu
\tilde{q}_\nu}{4t^2}\right)
+(1-2 x)^2 q_\mu q_\nu-\right.\nonumber\\
&&\left. -2 \left(\frac{d}{2t}-\frac{|\tilde{q}|^2}{4t^2}+x^2
q^2\right) g_{\mu\nu}\right]
e^{-|\tilde{q}|^2/4t-t x(1-x) q^2 }=\nonumber\\
&=&\frac{1}{2}\frac{1}{(4\pi)^{d/2}}\int_0^1 dx\int_0^\infty dt \,
t^{1-d/2} \left[g_{\mu\nu} \left(\frac{2-d}{t}+\frac{|\tilde{q}|^2}{2t^2}-2x^2 q^2\right)\right.\nonumber\\
&&\left. +(1-2 x)^2 q_\mu q_\nu
-\frac{\tilde{q}_\mu\tilde{q}_\nu}{t^2}\right]
e^{-|\tilde{q}|^2/4t-t x(1-x) q^2 }. \eea Now the coefficient of
$g_{\mu\nu}$ can be rearranged with the help of the following
identity \bea && \left(\frac{|\tilde{ q}|^2}{2t^2}-2x^2
q^2\right)\! e^{-\frac{|\tilde{ q}|^2}{4t}-t x(1-x) q^2 }=
\nonumber\\
\!\!&=&\!\! \left[2\frac{d}{dt}\left(-\frac{|\tilde{q}|^2}{4 t}-t
x(1-x) q^2\right)\!\!+
2 x(1-x) q^2-2x^2 q^2\right]\!e^{-\frac{|\tilde{q}|^2}{4t}-t x(1-x) q^2 }\!\!\!=\nonumber\\
\!\!&=&\!\!\left[2\frac{d}{dt}\left(-\frac{|\tilde{q}|^2}{4 t}-t
x(1-x) q^2\right)
-(2 x-1)^2 q^2+(1-2x)q^2\right]      e^{-\frac{|\tilde{q}|^2}{4t}-t x(1-x) q^2 }       \!\!\!=\\
\!\!&=&\!\!\left[2\frac{d}{dt}\left({e^{-\frac{|\tilde{q}|^2}{4t}-t
x(1-x) q^2}}
 \right)
-(2 x-1)^2 q^2{e^{-\frac{|\tilde{q}|^2}{4t}-t x(1-x) q^2}}
-\frac{1}{t}\frac{d}{dx}\left({e^{-\frac{|\tilde{q}|^2}{4t}-t
x(1-x) q^2}}
 \right)\right].\nonumber
\eea The last term vanishes when integrated over $x$ since  the
integrand in $x=0$ and $x=1$ takes the same value. The first term
instead can be rewritten as follows \bea \int^\infty_0 dt \,
t^{1-d/2}2\frac{d}{dt}\left({e^{-\frac{|\tilde{q}|^2}{4t}-t x(1-x)
q^2}}
 \right)=-(2-d)\int^\infty_0 dt t^{-d/2}\ e^{-\frac{|\tilde{q}|^2}{4t}-t x(1-x) q^2}.
\eea This contribution exactly cancels the similar contribution in
eq. (\ref{pi8}). Thus we are left with \bea
I^{\mathrm{n-pl.}}_{\mu\nu}\!\!&=&\!\!\frac{1}{2(4\pi)^{d/2}}\int_0^1
\!\!\!dx\!\!\int_0^\infty\!\!\!\! dt~ t^{1-d/2} \left[(q_\mu
q_\nu-g_{\mu\nu}q^2)(1-2x)^2  -\frac{\tilde{q}_\mu
\tilde{q}_\nu}{t^2}\right]
e^{-\frac{|\tilde{q}|^2}{4t}-t x(1-x) q^2 }=\nonumber\\
\!\!&=&\!\!\frac{1}{(4\pi)^{d/2}}\int_0^1\!\!\! dx \left[(q_\mu
q_\nu-g_{\mu\nu}q^2)(1-2x)^2\left(\frac{|\tilde{q}|}{2
M}\right)^{2-\frac{d}{2}}
  \!\!\!\!\!\!\!\!\!\Mfunction{K}_{\frac{d}{2}-2}\left(M |\tilde{q}|\right)-\right.\nonumber\\
  &&\left.
    -\frac{\tilde{q}_\mu \tilde{q}_\nu}{\tilde{q}^2}
 2 M^2\left(\frac{|\tilde{q}|}{2 M}\right)^{2-\frac{d}{2}}
\!\!\!\! \Mfunction{K}_{\frac{d}{2}}\left(M
|\tilde{q}|\right)\right]. \eea

\end{document}